\documentclass[sigconf, nonacm]{acmart}

\usepackage[utf8]{inputenc} 
\usepackage[T1]{fontenc}    
\usepackage{hyperref}       
\usepackage{url}            
\usepackage{booktabs}       
\usepackage{amsfonts}       
\usepackage{nicefrac}       
\usepackage{microtype}      
\usepackage{xcolor}         

\usepackage{amsmath}
\usepackage{mathtools}
\usepackage{float}
\usepackage{subcaption}
\usepackage{multirow}
\usepackage{adjustbox}
\usepackage{soul}
\usepackage{marvosym}
\usepackage{bbm}
\usepackage{siunitx}
\usepackage{lipsum}
\usepackage{tabularx}
\usepackage{algorithm}
\usepackage{enumitem}
\usepackage{wrapfig}
\usepackage{caption}  
\usepackage{pifont}

\usepackage{titletoc}

\usepackage{algorithm}
\usepackage{algpseudocode}
\usepackage{nicefrac}

\newcommand{\xmark}{\ding{55}}%

\definecolor{darkgreen}{rgb}{0.0, 0.7, 0.0}  
\definecolor{darkred}{rgb}{0.7, 0.0, 0.0} 

\newcolumntype{M}[1]{>{\centering\arraybackslash}m{#1}}

\DeclareMathOperator*{\argmin}{argmin}

\newcommand\vldbdoi{XX.XX/XXX.XX}
\newcommand\vldbpages{XXX-XXX}
\newcommand\vldbvolume{14}
\newcommand\vldbissue{1}
\newcommand\vldbyear{2020}
\newcommand\vldbauthors{\authors}
\newcommand\vldbtitle{\shorttitle} 
\newcommand\vldbavailabilityurl{URL_TO_YOUR_ARTIFACTS}
\newcommand\vldbpagestyle{plain} 

\begin{document}
\title{B+ANN: A Fast Billion-Scale Disk-based Nearest-Neighbor Index}

\author{Selim Furkan Tekin$^*$}\thanks{$^*$This work was done while the author was at IBM T. J. Watson Research Center.}
\affiliation{%
  \institution{Georgia Institute of Technology}
  \city{Atlanta}
  \state{Georgia}
}
\email{stekin6@gatech.edu}

\author{Rajesh Bordawekar}
\affiliation{%
  \institution{IBM T. J. Watson Research Center}
  \city{Yorktown Heights}
  \state{New York}
}
\email{bordaw@us.ibm.com}

\begin{abstract}
Storing and processing of embedding vectors by specialized Vector
databases (VDBs) has become the linchpin in building modern AI
pipelines. Most current VDBs employ variants of a graph-based
approximate nearest-neighbor (ANN) index algorithm, HNSW, to answer semantic
queries over stored vectors. Inspite of its wide-spread use, the HNSW
algorithm suffers from several issues: in-memory design and
implementation, random memory accesses leading to degradation in cache
behavior, limited acceleration scope due to fine-grained pairwise
computations, and support of only semantic similarity queries. In this
paper, we present a novel disk-based ANN index, B+ANN, to address these
issues: it first partitions input data into blocks containing
semantically similar items, then builds an B+ tree variant to store
blocks both in-memory and on disks, and finally, enables hybrid edge- and
block-based in-memory traversals. As demonstrated by our experimantal
evaluation, the proposed B+ANN disk-based index improves both quality
(Recall value), and execution performance (Queries per second/QPS)
over HNSW, by improving spatial and temporal
locality for semantic operations, reducing cache misses (19.23\%
relative gain), and decreasing the memory consumption and disk-based
build time by 24$x$ over the DiskANN algorithm. Finally, it enables
dissimilarity queries, which are not supported by similarity-oriented
ANN indices. 
\end{abstract}


\maketitle

\pagestyle{\vldbpagestyle}
\begingroup\small\noindent\raggedright\textbf{PVLDB Reference Format:}\\
\vldbauthors. \vldbtitle. PVLDB, \vldbvolume(\vldbissue): \vldbpages, \vldbyear.\\
\href{https://doi.org/\vldbdoi}{doi:\vldbdoi}
\endgroup
\begingroup
\renewcommand\thefootnote{}\footnote{\noindent
This work is licensed under the Creative Commons BY-NC-ND 4.0 International License. Visit \url{https://creativecommons.org/licenses/by-nc-nd/4.0/} to view a copy of this license. For any use beyond those covered by this license, obtain permission by emailing \href{mailto:info@vldb.org}{info@vldb.org}. Copyright is held by the owner/author(s). Publication rights licensed to the VLDB Endowment. \\
\raggedright Proceedings of the VLDB Endowment, Vol. \vldbvolume, No. \vldbissue\ %
ISSN 2150-8097. \\
\href{https://doi.org/\vldbdoi}{doi:\vldbdoi} \\
}\addtocounter{footnote}{-1}\endgroup

\ifdefempty{\vldbavailabilityurl}{}{
\vspace{.3cm}
\begingroup\small\noindent\raggedright\textbf{PVLDB Artifact Availability:}\\
The source code, data, and/or other artifacts have been made available at \url{\vldbavailabilityurl}.
\endgroup
}


\section{Introduction}

With the rise of deep learning architectures, pioneered by the
natural language processing (NLP) models, latent information from
unstructured~\cite{mikolov:corr-abs-1301-3781, vaswani2023attentionneed} and
structured data~\cite{bordawekar:corr-abs-1603-07185,bordawekar:deem2017} can now be encoded into high-dimensional vector
representations known as
\emph{embeddings}. The semantic relationships
among such diverse types of data can then be quantitatively captured via distance
metrics (e.g., euclidean or cosine similarity) computed over their
embeddings. This has revolutionized the storage and processing of
information and triggered the emergence of a new class of data
management systems,  termed as Vector databases (VDBs). VDBs 
are being widely adopted for storing and processing semantic vector embedding representations of large-scale structured
and unstructured content\cite{wang2021milvus, elastic, chroma,
  yang2020pase, sqldi, db2vectors, oracle26i}. The interest in exploiting VDBs grew enormously with the
collaboration of VDBs and Large Language Model (LLM) pipelines, such
as those built around models like OpenAI GPT, Google Gemini, Anthropic
Claude, and others. These pipelines generate responses with the retrieved context
gathered by the similarity search in the VDBs, which is known as
Retrieval Augmented Generation (RAG)~\cite{lewis2020retrieval}. The
wide-spread use of VDBs has motivated researchers to revisit the core
technical feature of VDBs: \textit{similarity search} index
algorithms. Essentially, given a similarity query, each VDB retrieves the top-$k$ nearest
embeddings for that query using an index. Since calculating the exact Nearest
Neighbors (NN) takes $O(N^2)$ complexity, approximate nearest neighbor
(ANN) indexing algorithms are used to decrease the search time up to
$O(\log N)$ at the cost of accuracy. The speed of algorithms depends
on two factors: (i) the number of distance computations to reach the
final answer, and (ii) the number of memory and disk retrievals
performed before reaching a sufficiently similar vector to the query
in the database. Even though there is a tremendous amount of activity
for targeting the first factor, most algorithms assume the data will be
stored in memory and disregard factor (ii) and the locality of stored
data.

\begin{table}[t]
  \begin{adjustbox}{width=0.49\textwidth, center}
    \centering
    \begin{tabular}{p{1.9cm} M{1.2cm} M{1.2cm} M{1.2cm} M{1.2cm} M{1.2cm} M{1.2cm} M{1.2cm} }
        \toprule
        Method & On Disk & Hybrid-Mem. Search & Dissimi-larity S. & Memory Cap & View Creation & Async. Update & Arch. Agnostic \\
        \hline
        HNSW \cite{malkov2014approximate} & \textcolor{darkred}{\xmark} & \textcolor{darkred}{\xmark} & \textcolor{darkred}{\xmark} & \textcolor{darkred}{\xmark} & \textcolor{darkred}{\xmark} & \textcolor{darkred}{\xmark} & \textcolor{darkgreen}{\checkmark}\\
        DISKANN \cite{jayaram2019diskann} & \textcolor{darkgreen}{\checkmark} & \textcolor{darkred}{\xmark} & \textcolor{darkred}{\xmark} & \textcolor{darkred}{\xmark} & \textcolor{darkred}{\xmark} & \textcolor{darkred}{\xmark} & \textcolor{darkred}{\xmark} \\
        SPTAG \cite{ChenW18} & \textcolor{darkgreen}{\checkmark} & \textcolor{darkred}{\xmark} & \textcolor{darkred}{\xmark} & \textcolor{darkred}{\xmark} & \textcolor{darkred}{\xmark} & \textcolor{darkred}{\xmark} & \textcolor{darkred}{\xmark} \\
        SCANN \cite{guo2020accelerating} & \textcolor{darkred}{\xmark} & \textcolor{darkred}{\xmark} & \textcolor{darkred}{\xmark} & \textcolor{darkred}{\xmark} & \textcolor{darkred}{\xmark} & \textcolor{darkred}{\xmark} & \textcolor{darkred}{\xmark} \\
        \hline
        \textbf{B+ANN} & \textcolor{darkgreen}{\checkmark} & \textcolor{darkgreen}{\checkmark} & \textcolor{darkgreen}{\checkmark} & \textcolor{darkgreen}{\checkmark} & \textcolor{darkgreen}{\checkmark} & \textcolor{darkgreen}{\checkmark} & \textcolor{darkgreen}{\checkmark}  \\
         \bottomrule
    \end{tabular}
    \end{adjustbox}
    \caption{Functionality Comparison of widely used ANN algorithms and B+ANN.}
    \label{tab:functionality_comparison}
\end{table}

Combined with the proliferation of unstructured data and the growing
interest in creating semantic relations in real-time, applications
require VDBs that can store petabytes of data and run similarity search
algorithms within seconds. Therefore, the assumption of adequate
memory for searching is untenable, rendering in-memory algorithms
impractical. Recent works such as \cite{jayaram2019diskann,
  chen2021spann, ChenW18, shim:vldb-ssd} have proposed disk-based solutions that
leverage SSDs and \texttt{mmap()} system calls. They adopt graph-based
solutions inspired by the Hierarchical Navigable Small Worlds (HNSW)
\cite{malkov2014approximate}, which is the most popular ANN indexing
algorithm. However, as demonstrated in this paper, HNSW, an \emph{in-memory} ANN index, performs
numerous random memory accesses, resulting in a high rate of cache
misses, and at every step, performs small numerical computations that
are difficult to optimize. Transitioning to the disk-based implementation exacerbates the
problem as these algorithms rely on \texttt{mmap()}-based access without
optimizing for memory locality. Furthermore, persisting the constructed
index to disk while maintaining high spatial locality is particularly
challenging for complex data structures such as graphs, where each
node must additionally store multiple edge connections. 

Alongside these, the VDBs are an emergent technology that lacks most
of the capabilities and system-based optimizations of traditional
relational databases. Most notably, while the temporal characteristics of subsequent queries is exploited in relational
databases, VDBs are still considered as a one-shot query
matching mechanism without temporal correlation. However, in a
conversation with a chatbot in a RAG system, users can ask temporarily
related questions as shown in Figure
\ref{fig:seq_motivation_b}. Considering the growth of agentic systems
and the pursuit of researchers to decrease time to first token
(TTFT) and inter-token latency (ITL), there is a high need
for a mechanism that exploits the temporal correlation for an
interactive experience with the user. In relational databases,
it is possible to make fine-grained access with designed access
control patterns by a buffer mechanism that maximize the cache
coherency; in contrast, VDBs provide only coarse-grained or no access
control mechanisms. Moreover, the modern relational database systems contains
many functionalities such as filtering and join operations not only
for similarity but also for dissimilar queries (e.g., Db2 z/OS SQL Data Insights
\texttt{AI\_SIMILARITY} and \texttt{AI\_COMMONALITY}
queries~\cite{sqldi}), whereas, most VDBs only provide most
similar vectors for a given query. The dissimilarity task is not
easily achievable, due to current indexing algorithms are based on graphs connecting
vertices with their most-similar nearest neighbor. Therefore, a dissimilar query
usually oscillates between two furthest vertices in the graph that are
highly unrelated.   

To this end, this paper targets a similarity search indexing algorithm for VDBs
that incorporates memory control, buffering, paging, persistence
management, cache efficiency, and the leveraging of temporal
correlations. We propose \texttt{B+ANN}, a billion-scale, disk-based,
and memory-efficient ANN indexing algorithm. As the first line of attack on the
indexing problem, we recursively partition the subspace by
hierarchical clustering to obtain many clusters and their
centroids. Then, we repurpose the widely used B+ Tree approach to the
vector domain via using the centroids as keys and vectors as
values. Our generated B+ANN tree has low complexity and can be stored on
the disk by maximising locality between semantic vectors, and provides
a fine-grained access control mechanism during traversal. Second, we
create skip-connections between leaves of the in-memory version of the
B+ANN tree to perform graph traversal for the most precise searches. The resulting tree
index converges in fewer hops, is more precise, faster, and easier to
parallelize by exploiting local data storage and improved memory
locality. The B+ANN algorithm coalesces the batched queries to increase compute granularity, while graph-based
approaches of VDBs perform pair-wise the distance computations that
are too fine-grained to be accelerated efficiently. Third, we
introduce \textit{semantic views} that can be extracted 
from the tree to represent subsets of information that create a
context for a given query. Subsequent queries are answered from the
query-specific view, which allows contextualized searches that are optimized for related
semantic queries. We show that the view can survive for up to a
thousand queries if they are given temporally correlated.  

Table~\ref{tab:functionality_comparison} compares and summarizes key
features of the B+ANN ANN index against a set of widely used indexing
algorithms. We now highlight key contributions from our work:

\begin{itemize}[itemsep=0pt, parsep=0pt, topsep=0pt, left=0pt]

\item The B+ANN is designed from ground-up to be a disk-based
  ANN index. The choice of a B+ tree inspired data structure, allows the
  index to use the same data structure for both disk-based and
  in-memory processing.
  \item The in-memory B+ANN tree enables \emph{hybrid} traversals over
  the tree nodes either using \emph{skip} edges to access individual nodes or
  via edges between individual vectors across tree nodes. 
\item By exploiting semantic relationships, we improve spatial locality
  by storing the semantically-close vectors in physically-close disk
  and in-memory locations. We demonstrate that our design improves both
quality and search performance while decreasing cache misses up to 19.23\% relative gain and the index built time by $24x$. 
\item Once the locality of the ANN index is improved, the overall
  search problem becomes compute-bound. Moreover, small pair-wise compute vector operations can be now batched 
  into coarse-grained matrix computations and can be accelerated using
  hardware accelerators such as SIMD and GPUs.   
\item We leverage our spatial-locality optimized design to enable
  faster response time and longer view survival time of the temporally
  correlated queries, such as those generated in Agentic scenarios.
\item Our design incorporates unique functionalities of recent 
  AI-enabled relational databases (e.g., supporting dissimilarity queries) that
  are largely unexplored in VDBs.  
\end{itemize}

The remainder of this paper is organized as follows. Section
\ref{sec:related_work} reviews the related work and highlights the key
differences between our approach and existing methods. Section
\ref{sec:preliminaries} presents the preliminaries and fundamental
concepts necessary to understand the proposed framework. Section
\ref{sec:problem} formalizes the problem setting and defines the main
objectives. Section \ref{sec:method} details our proposed methodology
and system design. Section \ref{sec:eval} reports the experimental
evaluation and performance analysis. Finally, Section
\ref{sec:conclusion} concludes the paper with summarizing remarks and
discussions on potential future directions. 


\section{Related Work}
\label{sec:related_work}

Similarity search algorithms have been studied for half century and
more \cite{pan2023survey}. Today, in the rise of AI, they have become
the cornerstones of the vector databases  \cite{han2023comprehensive}
such as Milvus \cite{wang2021milvus}, Elasticsearch \cite{elastic},
Analytics-DB \cite{yang2020pase}, ChromaDB \cite{chroma}, and PASE
\cite{yang2020pase}. The popularity of the vector databases has grown
significantly due to the use of Maximum Inner Product Search (MIPS) to
retrieve necessary context for a query to a LLM, which is known as
Retrieval Augmented Generation (RAG) \cite{lewis2020retrieval}. Due to
their referencing property, vector databases are becoming increasingly
prevalent systems alongside LLMs. This has also led to the development
of hybrid queries containing attribute predicates such as giving a
date range together with \textit{top-k} searches \cite{zhang2023vbase,
  gollapudi2023filtered}.   

There have been numerous approaches to ANNs in the literature mainly
focus on the high-recall search with high throughput and low
latency. The approaches follow two main ideas: First, partitioning the
vector space into many sub-spaces to reduce the search space
iteratively, and second, creating proximity-based graphs and
navigating the closest node based on the query. In the partition-based
approaches, the space is dived by following clustering, hash, or tree
based solutions.  

Representative clustering based solutions, such as,
\cite{inverted_index, baranchuk2018revisiting} utilize inverted
indices (IVF) by dividing the space with clustering algorithms such as
K-means, and assigning a centroid as an index to represent all the
vectors in the subspace. The quantization based approaches, such as
Product Quantization (PQ) \cite{product_quantization} further divide
the input queries and the vectors by slicing into $M$ subspaces and
encoding them into codes to reduce the cost of storage
\cite{source_coding, wang2018composite}. Two consecutive works
\cite{guo2020accelerating, sun2024soar} further improved the
quantization performance by dynamic weighting on the reconstruction
loss based on the query. \citet{johnson2019billion} employ both PQ and
IVF, called as IVFPQ, to index billion scale data. However, IVF can
cause imbalanced partition on the data where some subspaces can be
overly populated. 

The hash-based solutions such as locality sensitive hashing (LSH)
\cite{datar2004locality, srp-charikar, andoni2008near, kalantidis2014locally,
  andoni2015practical}, maps the close vectors into the same buckets
and reach the candidate vectors that falls the same bucket with the
query. The algorithm works in sub-linear time and have theoretical
guarantees on the query accuracy \cite{jafari2021survey,
  learn_to_hash}. However, the hash table can grow significantly and
they are outperformed by more recent graph-based approaches.  

The tree-based solutions follow classical data structures to index the
vectors such as KD-Tree \cite{friedman1977algorithm},
R-Tree\cite{guttman1984r}, and M-Tree \cite{ciaccia1997m}, however,
their performance are limited with the number of dimensions. For
example, the KD-tree needs to alternate in every dimension to find the
closest leaf by traversing one-dimension at a time. Thus,
\citet{muja2014scalable} proposed high-dimensional tree-based solution
by creating an hierarchical tree with the recursive K-means
clustering. The constructed tree searches the query based on priority
queue sorted by distance to the next level nodes. Similarly,
\citet{annoy} proposed hierarchical tree constructed by dividing the
space with hyperplanes passing between two randomly selected points in
each recursive call. However, in both methods, the resulted trees can
be very deep and unbalanced, in addition, the insertions to the trees
are ignored. 

Instead of partitioning the space, the graph-based solutions, such as
HNSW, start with an empty graph and add each vector as a vertex. The
vertices are connected based on the distances to other vertices
\cite{malkov2014approximate}. The connectivity of the graph is an
hyper parameter where high connectivity creates less error but high
latency. To increase the search speed and escape from dense hubs, the
constructed graphs can contain skip connections
\cite{malkov2018efficient} or shortcuts across different levels of
HNSW graph~\cite{gong:shortcut}. MARGO~\cite{yue:vldb-margo} proposes
a monotonic path-aware graph layout on disks. Recently, more hybrid approaches such
as \cite{munoz2019hierarchical, jayaram2019diskann} are introduced by
utilizing hierarchical clustering and creating a graph in each
partition with strict connectivity. Then, the created graphs are
merged to create one final graph with strong connectivity with a
larger search space. These methods also utilize in-search updates to
perform pruning on the edges. However, these updates cause per-vertex
locks in parallel setup. The most recent approach
\cite{manohar2024parlayann} proposed a better scaling parallelization
of these algorithms with the generous assumption of memory
($>1\mathrm{TB}$). In addition to high memory cost, they start from
random seed node in each search operation resulting lots of cache
misses with random memory accesses.  

Recent works focusing on disk utilization (SSDs) and
being able to scale in billion with more modest memory assumption
($64-128\mathrm{GB}$) include DiskANN \cite{jayaram2019diskann},
HM-ANN \cite{ren2020hm}, and SPANN \cite{chen2021spann}. DiskANN
stores the large data such as full-precision vectors and the graph on
the SSD, while it uses  the compressed vectors during search time to
traverse on the graph. However, the final ranking of the candidate
vectors uses the full-precision vectors, and it also suffers from the
latency caused by the random disk accesses. HM-ANN divides the memory
into two by placing pivot points in fast and NSW graphs in to slow
memory. Still, it ends-up causing 1.5 times more fast memory
consumption. Lastly, SPANN stores the uncompressed posting lists in
disk (SSDs) and centroids in memory. It partitions the space by
following an hierarchical balanced clustering. However, the hierarchy
of the created tree is not fully utilized, i.e. only the leaf nodes
are used. The algorithm also requires multiple in-search update steps
to keep the tree balanced. 

Furthermore, utilization of GPUs into the ANN distance calculations
prevail among more recent systems such as FAISS \cite{douze2024faiss},
SONG \cite{zhao2020song}  and CAGRA \cite{ootomo2023cagra} (available in the Nvidia cuVS~\cite{cuvs} library). Compared
to FAISS, SONG and CAGRA accelerate graph-based
ANN on Nvidia GPUs, by exploiting advanced Nvidia CUDA features such as the \emph{warp} functions.

\section{Preliminaries}
\label{sec:preliminaries}


\begin{figure}[htbp]
  \centering
  \begin{subfigure}[b]{0.35\textwidth}
    \centering
    \includegraphics[width=\textwidth]{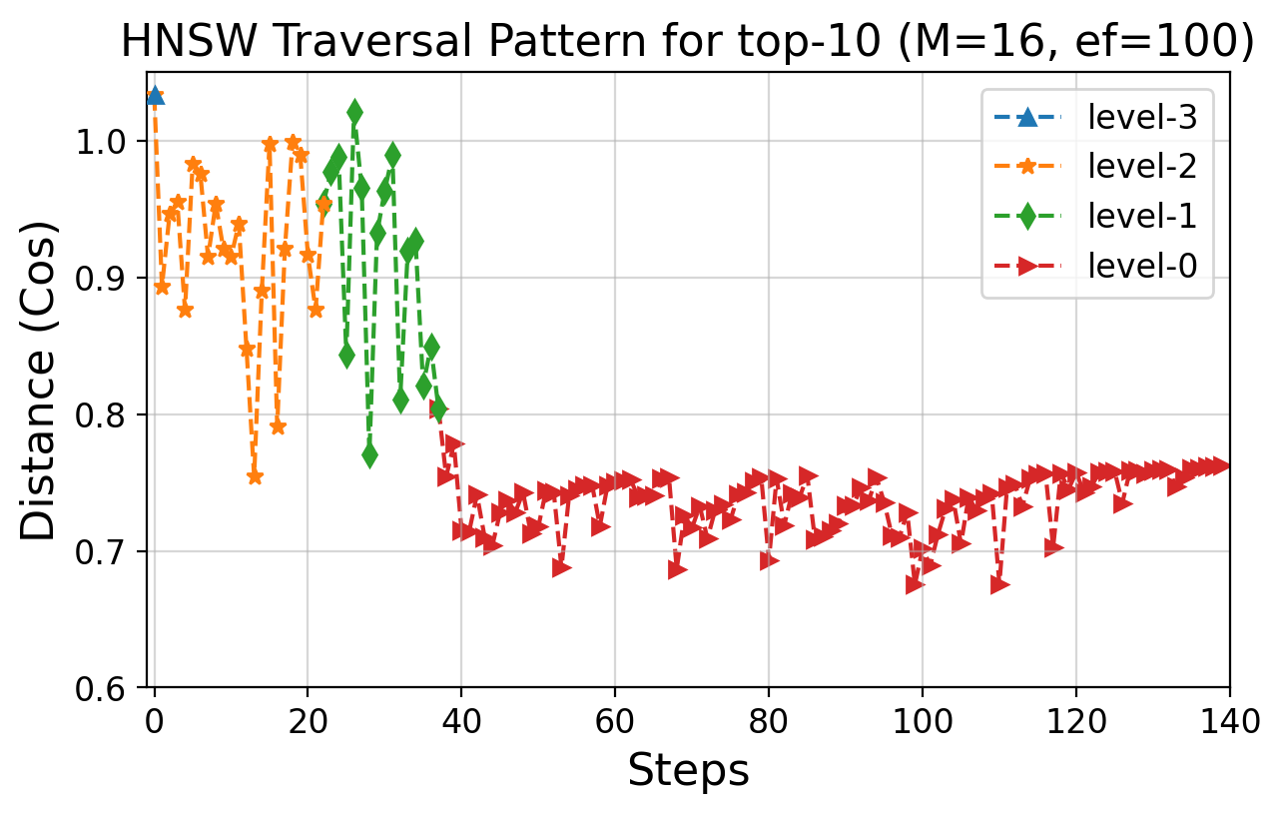}
    \caption{}
    \label{fig:hnsw_motivation_a}
  \end{subfigure}
  \hfill
  \begin{subfigure}[b]{0.35\textwidth}
    \centering
    \includegraphics[width=\textwidth]{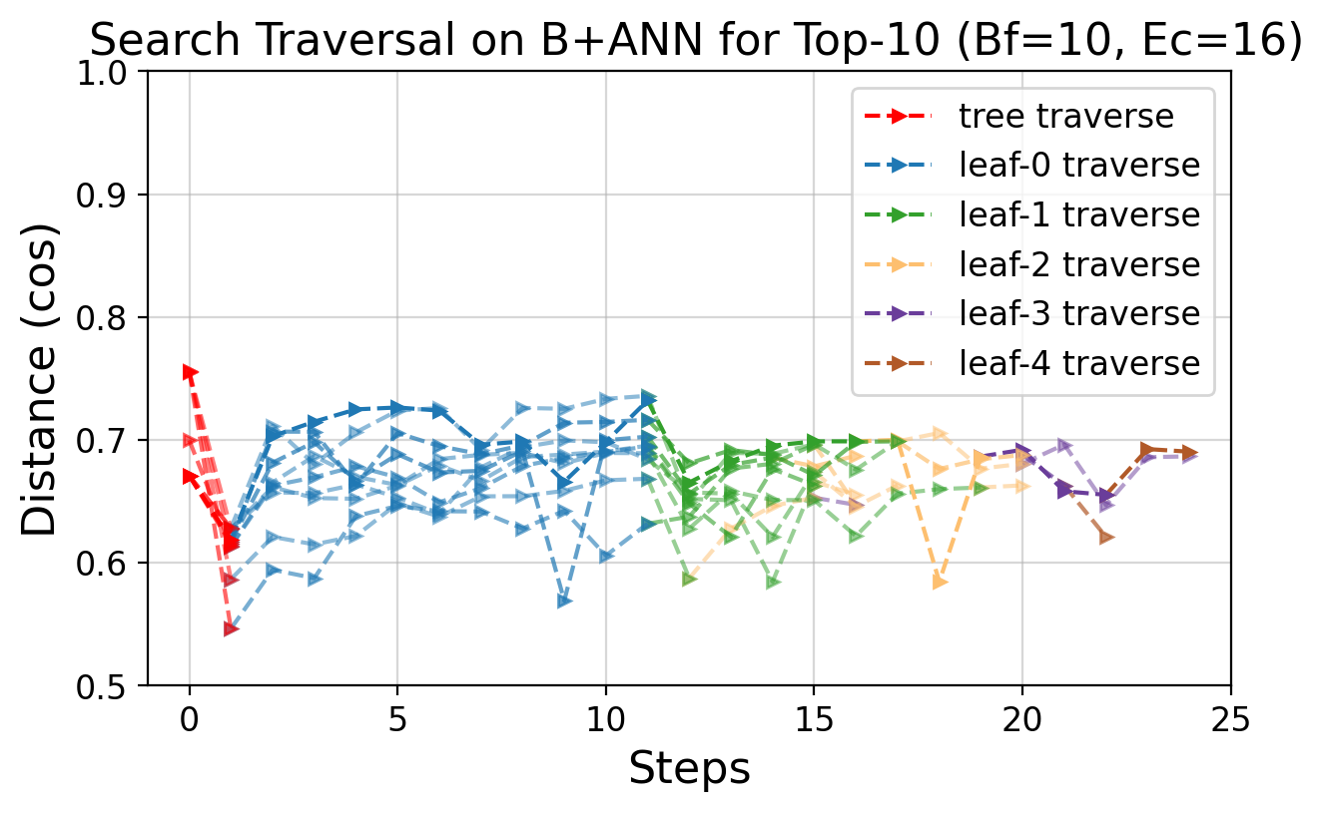}
    \caption{}
    \label{fig:hnsw_motivation_b}
  \end{subfigure}
  \caption{(a) The figure illustrates the HNSW retrieval pattern from
    the highest to the lowest level. Each visit to a node involves
    accessing memory and performing a pairwise distance
    calculation. (b) The retrieving pattern of B+ANN. The first phase
    is the tree traverse, and the second phase is the skip-edge
    connections. Each leaf node incurs one memory access, followed by
    a vector–matrix operation that computes the distances between the
    query and all vectors stored in the leaf. Note that the number of B+ANN
    steps (23) is substantially lower than HNSW (140).}  
  \label{fig:mainfig}
\end{figure}

\subsection{Random Accesses of HNSW}

The growing popularity of vector databases leads to increased
utilization of ANN algorithms, which enable fast similarity search
while balancing accuracy and computational efficiency. HNSW, in this
matter, is ubiquitous for the most popular VDBs, e.g.,
Chroma, Milvus, and pgVector, which is an open-source PostGRES
extension that supports vector data types and enables ANN search,
including HNSW. The reasons for the widespread use of HNSW are that it
offers flexibility and performance with efficient updates, without
degradation in recall performance. However, as we demonstrate in this
section, HNSW is not cache-efficient, as it employs a greedy-based
traversal with random initiations, resulting in cache misses and
slower memory accesses that exacerbate in disk-resident executions.  

Figure~\ref{fig:hnsw_motivation_a} presents an empirical
proof of these problems. HNSW starts at a random location at the
highest layer and traverses to the lower layers by jumping between the
nodes until it reaches the closest vicinity of the query vector it can
find. First, every visit to a node is an access to the memory, where
the nodes are located randomly in memory. Second, in each visit, the
greedy search calculates the distance between the query and the node’s
edges in a fine-grained manner through pairwise distance
computation. As an alternative, we propose a more efficient solution
to these two aspects by a \emph{hybrid} design of combining the B+Tree
structure with the modified greedy-search between leaf nodes, similar
to HNSW. B+ANN preserves node locality by placing semantically related
vectors in close physical proximity within memory or disk, and further
improves cache coherence. Second, as illustrated in
Figure~\ref{fig:hnsw_motivation_b}, a leaf-node in the B+ANN contains
multiple vectors, which allows batch loading and faster distance
calculation operations over multiple vectors, e.g., using matrix
multiplication instead of pair-wise dot-products. Therefore, with one
I/O operation, we can load many close vectors to the query and calculate
distances for more vectors at a time, and during the traversal, the
next leaf location is in proximity, thereby reducing access time.

\subsection{The Temporal Correlation in Retrievals}


\begin{figure}[htbp]
  \centering
  \begin{subfigure}[b]{0.30\textwidth}
    \centering
    \includegraphics[width=\textwidth]{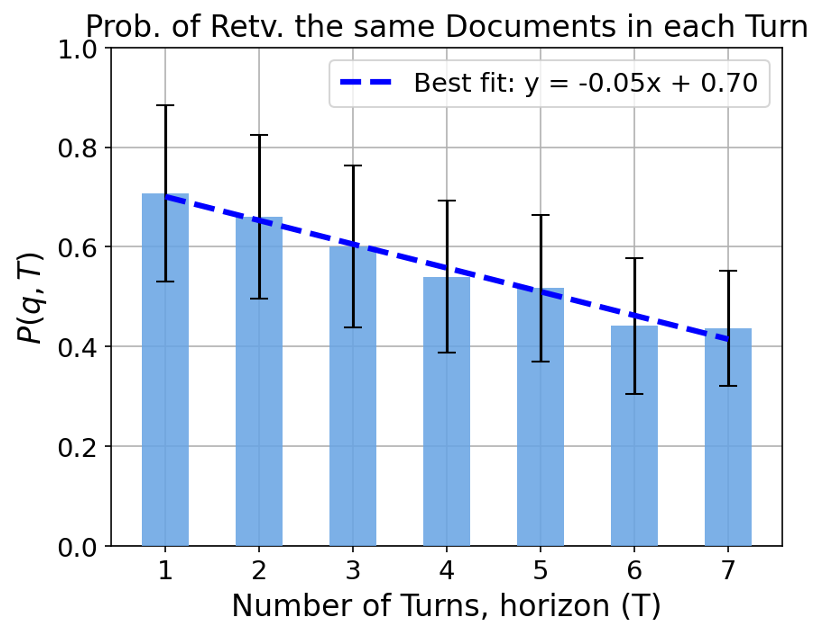}
    \caption{}
    \label{fig:seq_motivation_a}
  \end{subfigure}
  \hfill
  \begin{subfigure}[b]{0.30\textwidth}
    \centering
    \includegraphics[width=\textwidth]{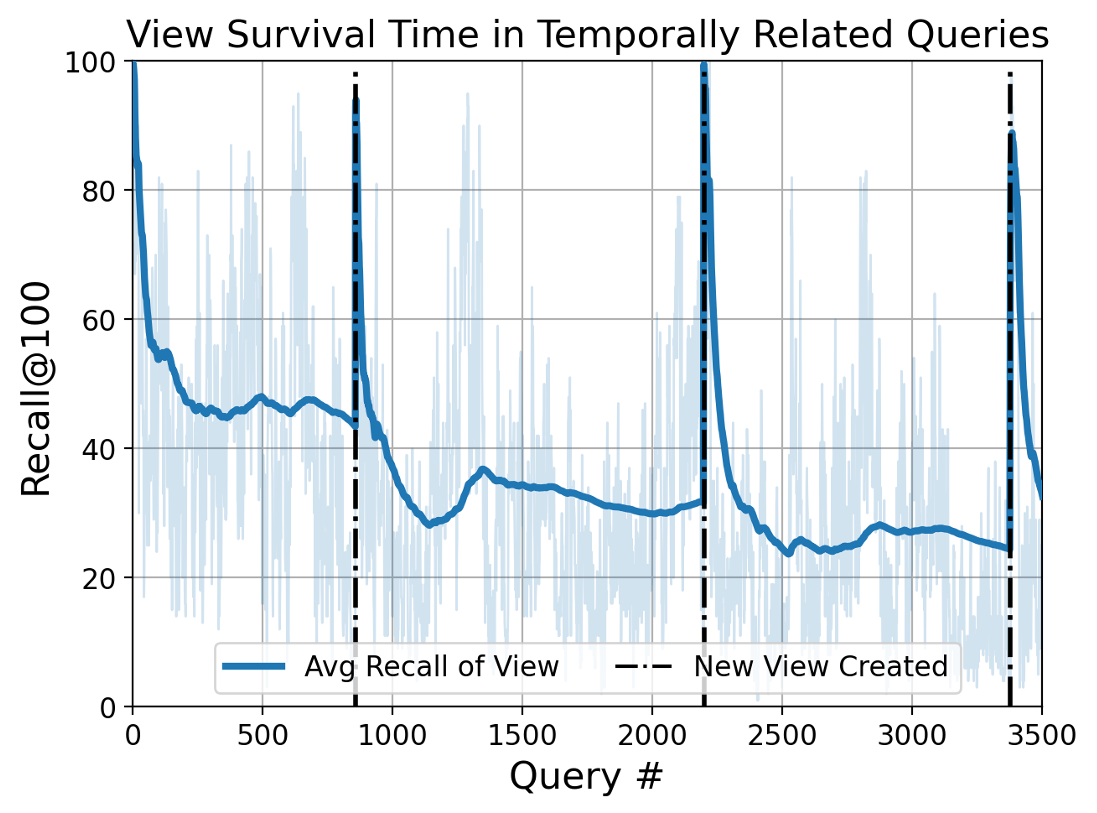}
    \caption{}
    \label{fig:seq_motivation_b}
  \end{subfigure}
  
  \caption{(a) We show our observation in multi-turn conversation of a RAG system \cite{katsis2025mtrag}: The probability of retrieving the same document in a conversation with an LLM for each turn. (b) Our proposed view creation system which exploits the temporal relation of successive queries.}
  \label{fig:mainfig2}
\end{figure}

Consider a Retrieval Augmented Generation (RAG) system and user
interaction in a multi-turn conversation. It is highly probable that
the first question asked by the user will be related to the following
questions. On this basis, it is also expected that the retrieved
information from a database for questions will overlap and
intersect. Therefore, there is a temporal relation between the vectors
retrieved in each turn of conversation. The left plot in Figure
\ref{fig:seq_motivation_a} shows an experiment on a multi-turn
conversation dataset of a RAG system, MTRAG \cite{katsis2025mtrag},
providing experimental evidence. As the number of turns increases, the
probability of retrieving the same documents decreases from 0.7 to
0.4, showing that there is a high temporal relationship between
consecutive queries. In light of this observation, we design a
modified B+Tree structure and introduce the \texttt{create\_view}
function that puts a granular perspective of a subset of the database
into the memory and responds in-memory (see Section ~/ref{sec:buildtree}for more
details). The right plot on Figure \ref{fig:seq_motivation_b} shows
how long a \texttt{view} can respond to temporally correlated queries
until the asked query is not found in the view. The spikes show that
the recall of the queries are highest once the view is created, and it
degrades as more queries come. 

A key functionality of a B+Tree structure is the sequential access
support by traversing the leaf nodes for range queries. In our
modified B+Tree structure, where the keys are centroids of clusters
and values are the vectors stored in those clusters, we created
\textit{skip connections} between leaf nodes (see
Section~\ref{sec:skip-conn} for more details). If the incoming queries are temporally correlated, the
subsequent vectors are likely to reside in the same node or in
neighboring nodes. By traversing adjacent leaf nodes, these vectors
can be retrieved with improved memory access efficiency and cache
locality. Therefore, B+ANN excels at temporally correlated queries by
mapping temporal relations to the  spatial localities of
vectors. B+ANN is not limited with these functionalities as shown in
Table \ref{tab:functionality_comparison}. In the following sections,
we talk about methodology and other features of our design. We begin
by describing the ANN problem and defining our essential elements at
our formulation.

\section{Problem Formulation}
\label{sec:problem}

For a vector collection $\mathcal{D} = \{\mathbf{x}_i\}_{i=0}^{N}$ of $N$ vectors, where each \(\mathbf{x}_i \in \mathbb{R}^n\) represents the extracted semantic information, e.g., embedding, of a sample in the database with dimensionality of $n$. A distance function $d: \mathbb{R}^n \times \mathbb{R}^n \rightarrow \mathbb{R}$ 
maps vectors \(\mathbf{x}_i, \mathbf{x}_j\), where \(i,j = 1, \dots, N\) onto a scalar distance $d(\mathbf{x}_i, \mathbf{x}_j) = s$. Multiple distance functions can be used, e.g., Hamming, Minkowski, Mahalanobis but most commonly Euclidean distance ($L_2$ norm), and cosine distance ($1 - \cos(\theta)$) are used. Larger distances indicate more dissimilar input vectors, and zero indicates identical vectors. For the search of the most similar vectors, we define the following elements:

\textbf{Definition-1:} ($k$-\emph{Nearest Neighbors}) ($k$-NNs) Given a vector collection \(\mathcal{D}\) and a query \(\mathbf{q}\) in the same Euclidean space \(\mathbb{R}^n\), NN aims to find \(k\) nearest neighbors \(\mathcal{R}\) of \(\mathbf{q}\) by calculating a distance \(d(\mathbf{x},\mathbf{q})\) for each \(\mathbf{x} \in \mathcal{R}\), i.e.,
\begin{equation}
R(\mathbf{q}, k) = \argmin_{\mathcal{R} \subset \mathcal{D}, |\mathcal{R}| = k} \sum_{\mathbf{x} \in \mathcal{R}} d(\mathbf{x},\mathbf{q}).
\end{equation}
Here $R: Q \rightrightarrows \binom{\mathcal{D}}{k}$ is the set-valued map from query set $Q$ to the set of $k$-subsets of $\mathcal{D}$. The time complexity of the brute-force $k$-NN algorithm is \(O(k \cdot N \cdot n)\). For large vector collections (\(|\mathcal{D}| \gg \text{billions}\)), brute-force NNs are highly impractical. Therefore, researchers relaxed the definition of NNs and introduced \emph{Approximate Nearest Neighbor} (ANNS) algorithms, allowing more practical implementations with efficiency-accuracy trade-offs. 

\textbf{Definition-2:} ($k$-ANN) ANNS builds an index \(\mathcal{I}\) over the dataset \(\mathcal{D}\), which maps the input query $\mathbf{q}\in\mathbb{R}^{n}$ to a subset \(\mathcal{C} \subseteq \mathcal{D}\). At each step of propagation on the \(\mathcal{I}\), it computes distances \(d(\mathbf{x},\mathbf{q})\) between query and vectors \(\mathbf{x} \in \mathcal{C}\) to obtain the approximate \(k\) nearest neighbors \(\widetilde{\mathcal{R}}\) of \(\mathbf{q}\). The result is sufficiently close to the true set $\mathcal{R}$ with the gain in efficiency. To measure the level of correctness, recall at a subset size defined as follows:

\textbf{Definition-3:} (\textit{Recall}-$k@k'$) Let $\widetilde{\mathcal{R}}\subset\mathcal{D}$ be the output of ANN index $\mathcal{I}$ with size $k'$ for a query $\mathbf{q}$ and $\mathcal{R}$ be the true $k$-NNs of $\mathbf{q}$. Then the \textit{Recall}-$k@k'$ of $\mathbf{q}$ is $\frac{|\widetilde{\mathcal{R}} \cap \mathcal{R}|}{|\mathcal{R}|}$. The most common choice of recall is $10@10$, yet recently, in RAG systems, it is more common to use broad-retrieval and perform re-ranking among the retrieved top-$100$ vectors using re-ranking models \cite{rerankGAO, an2025hyperrag, chen2024bge}. In this paper, therefore, we focused on the recall $10@10$ and $100@100$.

The index of B+ANN, $\mathcal{I}$, is a combination of Tree-based and Graph-based indexing due to the skip-edge connections. The Tree-based structure partitions \(\mathcal{D}\) by making a small number of comparisons while still reaching sufficiently close vectors to the input query $\mathbf{q}$. Then, the Graph-Based structure, created by skip-connections, follows edges that lead to closer points in the graph with varying hop sizes. Based on these, we define the Tree and Graph indexing as follows:

\textbf{Definition-4:} (\textit{Tree-based} Indexing)
Given a finite dataset $\mathcal{D}$ in $\mathbb{R}^{n}$, a rooted tree denoted by $\mathcal{T} = (\mathcal{V}, \mathcal{E}, r)$ constructed on $\mathcal{D}$ as index $\mathcal{T}(\mathbf{q}, k)$ that returns $k$-ANNs. It consists of a set of vertices $\mathcal{V}$, a set of directed edges $\mathcal{E} \subseteq \mathcal{V}\times \mathcal{V}$, and a root node $r\in \mathcal{V}$. For any vertex, there is only one vertex that is connected to it, $\forall \upsilon\in \mathcal{V}\setminus\{r\}$, we have $\exists! u\in \mathcal{V},\; (\upsilon, u)\in \mathcal{E}$. To represent the hierarchy, each branch of the tree can be thought of as a subtree; then we can also define $\mathcal{T}= \{r, [\mathcal{T}_1, \mathcal{T}_2, \dots, \mathcal{T}_k]\}$ for a rooted tree with $k$ branches. We define the parent-child relation as $parent(\upsilon) = u\; \mathrm{if}\; (u, \upsilon)\in \mathcal{E}$ and $children(\upsilon) = \{u \mid (\upsilon, u)\in \mathcal{E}\}$. Therefore, the set of leaf nodes of $\mathcal{T}$ is denoted as $\mathcal{L}(\mathcal{T}) = \{\upsilon\in\mathcal{V} \mid children(\upsilon)=\emptyset \}$. Tree-based ANN reaches the $\widetilde{\mathcal{R}}$ by traversing from the root toward the leaves, selecting at each step the branch whose subspace (or cluster) is closest to the query. Each selected branch represents a hierarchical partition of the data space where the children of the selected parent reside. Once the lowest level (leaves) is reached, the search terminates by returning the top-$k$ closest vertices.

\textbf{Definition-5:} (\textit{Graph-based} Indexing) Given a dataset $\mathcal{D}$, a graph denoted by $G=(\mathcal{V, \mathcal{E}})$ is constructed as index $G(\mathbf{q}, k)$ that returns $k$-ANNs. Here, $V$ is the set of vertices denoted by $\upsilon, u\in\mathcal{V}$ and have neighborhood relationship $(u, \upsilon)\in \mathcal{E}$. For a query $\mathbf{q}$, seeds $\widetilde{\mathcal{C}}$, routing strategy, and terminal condition, $G$ conducts search starting from $\widetilde{\mathcal{C}}$ and continuously find subsets $\mathcal{C}\subseteq\mathcal{D}$ via routing strategy until $\mathcal{C}$ satisfies the terminal condition. The algorithm returns nearest vectors in $\mathcal{C}$ as approximate $k$-NNs $\widetilde{\mathcal{R}}$.

\textbf{Definition-6:} (\textit{View}) For a given dataset $\mathcal{D}$, the tree-based index $\mathcal{T}$, a query $q_t$ at time $t$, and size of view $k$, a view is extracted from the $\mathcal{T}$ by $k$-ANNs of $\mathbf{q}$ to create a sub-tree $\mathcal{T}'$, which is denoted as follows: 
\begin{equation}
\begin{split}
    V(q_t, \mathcal{T}, k) &= \mathcal{T}',\\
    \mathrm{s.t.}\; \mathcal{T}'=(\mathcal{V}', \mathcal{E}', r'),\; & \upsilon\in\mathcal{T}(q_t, k)\; \mathrm{and}\; \upsilon\in\mathcal{V}'\\
\end{split}
\end{equation}
where, $V$ denotes the view function that constructs a single tree $\mathcal{T}'$ from the set of $k$-ANNs of the query $\mathbf{q}$, retrieved by the base index $\mathcal{T}(q_t, k)$. Consequently, $\mathcal{T}'$ represents a local tree structure induced by the $k$ retrieved vectors that approximate the neighborhood of $\mathbf{q}$. For the upcoming queries $q_{>t}$, the sub-tree $\mathcal{T}'$ index and returns the $k$-ANNs.

Building on these definitions, next, we show our methodology to build B+ANN, how it indexes the queries, and unique functionalities to make optimum accesses with the least memory footprint.




\section{Methodology}
\label{sec:method}

\begin{figure*}[hbt!]
    \centering
    \includegraphics[width=\textwidth]{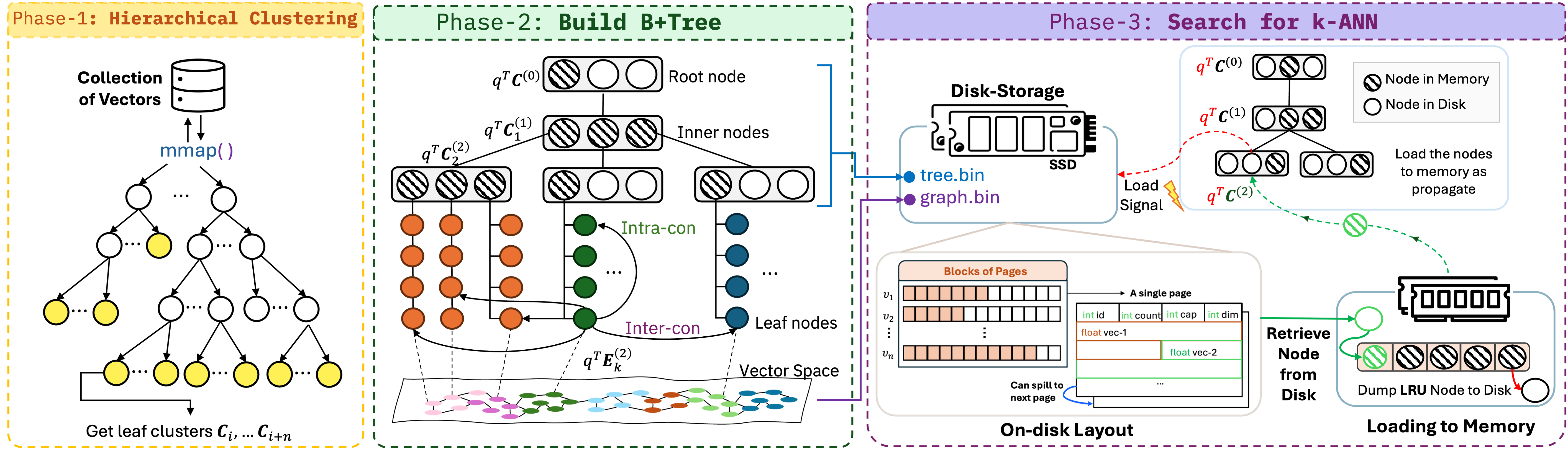}  
    \caption{We show the three phases of B+ANN Indexing. First phase partitions the vector space with hierarchical clustering. The second phase builds the B+ANN tree with skip-edge connections. The third phase indexes the query and performs accesses.}
    \label{fig:main_fig}
\end{figure*}

Based on the motivations presented in Section \ref{sec:preliminaries}, Figure \ref{fig:main_fig} illustrates the three phases of our design, B+ANN indexing. In this section, we cover the details of each phase, building upon the definitions provided in Section \ref{sec:problem}.

\subsection{Recursively Partitioning by Hierarchical Clustering}
\label{sec:hier_tree}

Considering the data of a large dataset ($|\mathcal{D}|\gg \text{billions}$) covering the space, it is common to observe non-uniform data distributions forming local hubs with thousands of vectors close to each other. When partitioning the space into subregions, the partitioning scheme should be adaptive to this distribution, allocating finer partitions to dense regions to better capture their local structure. To tackle the imbalanced partitioning, we perform recursive K-Means clustering. Specifically, we use K-means++ Clustering \cite{arthur2006k}, which reduces number of iteration compared to standard K-means and it finds better initial centroids by initializing them furthest to every other centroid. The algorithm is shown in Algorithm \ref{alg:sampling} which takes four parameters: dataset $\mathcal{D}$, number of clusters in each K-Means operation $K$, the threshold value $\tau$, which is the maximum number of vectors exists in a cluster, and $J$ is the number of expectations-maximization steps that each K-Means run. 

\begin{algorithm}[hbt!]
\caption{Hierarchical Clustering}\label{alg:sampling}
\small
    \begin{algorithmic}[1]
    \Procedure{hcluster}{$\mathcal{D}$, $\tau$, $K$, $J$}
            \State $q=\mathrm{Queue}(),\; \mathrm{clusters} = [\;]$
            \State $q.\mathrm{push}(\mathcal{D})$ \Comment{Add the initial data to the queue}
            \While{$!q.\mathrm{empty}()$}
                \State $\mathrm{data\_part} \gets q.\mathrm{pop}()$
                \State $\mathrm{clustered\_data}\gets \texttt{Kmeans++}(\mathrm{data\_part}, K, J)$ 
                \For{$i = 1$ to $K$}
                    \If{$\mathrm{clustered\_data}[i].\mathrm{size}() > \tau$}
                        \State $\mathrm{clusters}.\mathrm{add}(\mathrm{data\_part})$
                    \Else
                        \State $q.\mathrm{push}(\mathrm{clustered\_data}[i])$
                    \EndIf
                \EndFor
            \EndWhile
            \State \Return $\mathrm{clusters}$ \Comment{Return Clusters}
    \EndProcedure
    \end{algorithmic}
\end{algorithm}

As shown in line 8, the algorithm adds the partitioned data into the queue since it has a size higher than the threshold $\tau$. The popped stack contains the next data to be partitioned into clusters, as shown in line 6. Each clustering operation on a data partition creates another level at the hierarchical tree, as shown in Phase-1 in Figure \ref{fig:main_fig}. Where the leaves contains the final clusters $\mathbf{C}_1,\dots,\mathbf{C}_m$ with cluster count $m$ and each cluster having less than $\tau$ vectors denoted by $\mathbf{C}_i = [\mathbf{x}_{i,1}, \dots, \mathbf{x}_{i,n'}]$, where $n'\leq\tau$ and $\mathbf{x}_{i,j}\in \mathcal{D}$. 

Considering the computational cost, a K-means clustering operation has $O(K\times N\times J)$ complexity without considering the dimensionality of the vectors. The height of hierarchical tree is $\log_{K}(N)$, thereby, total Hierarchical Clustering has complexity of $O(K\times N\times J \times \log_{K}(N))$. Another crucial point is the memory cost, as line 3 indicates, when the first $K$-Means++ algorithm is applied on the full dataset, the cost can drastically increase. Loading the dataset at once into the memory can cause an OOM error and is dangerous; therefore, we employ \texttt{mmap()} system-call to let the OS handle the memory management during the Hierarchical Clustering. The vectors remain accessed through their memory-mapped representation until the tree construction is completed. Lastly, during the creation of clusters, we keep track of the hierarchy to use in our B+ Tree-inspired index creation.

\subsection{Building and Indexing B+ANN Tree}
\label{sec:buildtree}

B+ Tree is prominent for single-column indexing in relational databases, and due to its flexibility, it grows wide with low height, resulting in $O(\log N)$ complexity for search/insert/delete operations and $O(N\log N)$ complexity to build. The data is stored in leaves, and the inner nodes are only used for navigation. This allows better control of the memory by storing the inner nodes in memory and leaves in disk. With a buffer system that loads leaves from disk to memory from a block-oriented storage, it allows fine-grained control over the data for memory management. A differentiating factor of the B+Tree indexing is its sequential and range query performance, since the data is stored in leaves, and it can traverse from one link to another without going to the higher layers. Overall, a significant amount of experience and development has been made in relational vector databases and particularly in B+Tree algorithms. As the motivations of vector and relational databases are analogous, thereby, we adapt the B+ Tree indexing to vectors by using cluster centroids $\mathbf{c}_i^{(l)}$ as keys and their vectors $\mathbf{v}_j$ as values. Here, a centroid is defined:
\begin{equation}
    \mathbf{c}_i^{(l)} = \frac{1}{m}\sum_{j=1}^{m}\mathbf{v}_{i,j}^{(l)}, \; \mathrm{and}\; \mathbf{C}_i^{(l)}=[\mathbf{v}_{i, 1}^{(l)}, \dots, \mathbf{v}_{i, m}^{(l)}]
\end{equation}
where $m$ is the size of the cluster, $l$ is the level of the cluster in tree, $\mathbf{v}_{i,j}^{(l)}$ is the vector belonging to cluster $i$ at level $l$ such that $\mathbf{v}_{j}\in\mathbf{C}_{i}^{(l)}$ and $j=1,\dots,m$. The key distinction between inner and leaf nodes lies in their contents: Inner nodes store the centroids of their child nodes as values, whereas leaf nodes store the centroids of the actual data clusters as keys and the original vectors as their associated values. We denote the key value pairs as $\mathbf{c}^{(l)}_{i}\mapsto\mathbf{C}_{i}^{(l)}$ representing the centroid of the node $i$ and its cluster matrix $\mathbf{C}_{i}$ containing the vectors at level $l$. For a B+ANN tree with a height of $L$, an inner node $i$ at level $L-1$ contains centroids $\mathbf{C}_i^{(L-1)} = [\mathbf{c}_1^{(L)}, \dots, \mathbf{c}_{m}^{(L)}]$ where each centroid $j$ maps to each cluster at level $L$, as denoted by $\mathbf{c}_j^{(L)}\mapsto\mathbf{C}_j^{(L)}$. In Phase 2 of Figure \ref{fig:main_fig}, the structure of the modified B+ Tree, which we call \texttt{B+ANN}, is shown with the root node, inner nodes, and leaves that contain the vectors in the vector space.

B+ANN Tree is a self-balancing tree that grows wide by adding clusters based on their centroids to the leaf nodes. For each added cluster to the leaf node, the centroid is added to the inner node that points to the leaf. The inner node and the leaf node update their clusters. Similar to a B+ Tree, it grows horizontally when the nodes are fully loaded (or above some threshold value), and the node splits into two, creating two new centroids that are added to the upper level inner node. The split performs K-Means with 2 clusters, where the centroids of the split nodes are updated. The capacity of the inner node is determined by the ${\kappa}_\mathrm{inner}$ parameter.

\textbf{Building the B+ANN:} In Section \ref{sec:hier_tree}, we obtain multiple clusters with their centroids representing the mean of residing vectors. Instead of adding each vector to the tree one by one, we exploit the hierarchy that was discovered by the previous step in Section \ref{sec:hier_tree}. We traverse the hierarchical structure, as illustrated in Figure \ref{fig:main_fig}, and record the number of vectors contained within each visited subtree. If the vector count is lower than the leaf-capacity parameter, ${\kappa}_\mathrm{leaf}$, then the leaves of the subtree are selected. The selected leaf clusters are highlighted in yellow. Then, we cut the leaf clusters and create the lowest-level nodes of the B+ANN by adding them to the lowest layer as pairs of centroid-clusters. In other words, after the hierarchical Clustering process, we have many pairs of $(\mathbf{c}_1, \mathbf{C}_{1}), ..., (\mathbf{c}_m, \mathbf{C}_{m})$ that we add to the B+ANN Tree, instead of adding each vector one by one to the tree. This way, we accelerate the building process, as well as keep the hierarchical relation between the clusters by respecting the geometry. For each insertion, the centroids are added to the inner (parent) node, starting from the root. Each inner node splits into two if it reaches the inner node capacity denoted by ${\kappa}_\mathrm{inner}$. The tree grows in balance, resulting in a wide tree with a lower height. Overall, the building algorithm takes two parameters ${\kappa}_{\mathrm{leaf}}$ and ${\kappa}_\mathrm{inner}$, and the pairs of centroid-clusters and returns the B+ANN tree. 

Once the tree is built, we can save it into to disk, under file named \texttt{tree.bin}. We show the storage pattern of the nodes in the Phase-3 of Figure \ref{fig:main_fig} by illustrating the blocks of pages and zooming to a single page structure. Each node is stored in blocks of pages containing the meta data, centroid, and the vectors stored in the node. A key design insight is that the vectors in the same node are stored together such that the vectors are both semantically and physically close. To maximize spatial locality, nodes that are adjacent in the tree are also stored in close proximity on disk. For a sequence of queries that are temporarily close, the proximal similarity in the storage enables frequent access to the same memory locations, consequently, less number of cache misses occur.

\textbf{Search in the B+ANN:} B+ANN retrieves the $k$-ANNs of a query $\mathbf{q}$ by performing Breadth-First Search (BFS) on the tree. As shown in Phase-2 of Figure \ref{fig:main_fig}, for a node $i$ in level $l$, we compute the distances to the lower-level nodes by computing $d(\mathbf{q}, \mathbf{c}),\; \forall \mathbf{c}\in \mathbf{C}^{(l)}_i$. If the selected distance metric is cosine distance, then the computation is vector-matrix multiplication and subtraction (denoted by $1 - \mathbf{q}^\top \mathbf{C}_{i}^{(l)}$), which can be accelerated by SGEMM operations by using libraries such as \texttt{cuBLAS}, \texttt{OpenBLAS}. Furthermore, in batched queries the operation can further be expedited by matrix-matrix multiplication (e.g., $\mathbf{Q}^{\top}\mathbf{C}_{i}^{(l)}$), which speeds up computations significantly using BLAS or GPU acceleration. This optimization is natural for tree- or cluster-based ANN methods, where you compare a batch of queries to centroids (structured subsets). Graph-based methods (like HNSW, NSG, etc.) typically perform fine-grained, pointer-chasing traversals that are less amenable to matrix-matrix parallelization due to their irregular memory access and dynamic traversal paths. 

The traversal is controlled by the priority-queue, which orders the next level nodes based on the distances from their centroids to the queries, prioritizing those with smaller distances by putting them at front. At the next level, we pop nodes from the front of the queue, with the number of nodes controlled by the branching factor parameter $\beta$. Each node popped from the queue can be processed in parallel by a thread, and in each thread, we compute the distances to the centroids of the lower nodes and push them to the queue. A lock governs the synchronization of push and pop operations. At the lowest layer, instead of calculating the distance to centroids, we calculate the distance to the data vectors, and return the $k$-ANNs.

A node, depending on its initialization, can be accessed either memory or disk. For a tree that is fully stored in the disk, when a query comes, we perform Disk I/O for every node and we load the accessed nodes into the memory during traversal. This system maintains a small memory footprint by loading only the traversed nodes to the memory which account for roughly $0.01\%$ of the all nodes in the whole tree. However, continuously allocating new nodes in memory may eventually lead to an out-of-memory (OOM) error. Therefore, we utilize Least Recently Used (LRU) cache policy. The loaded node stays in the memory until it is least recently used. We illustrated this process in Phase 3 of the Figure \ref{fig:main_fig}. The control of the LRU queue is run in an independent thread which is synchronized by a lock. The Algorithm \ref{alg:search} shows the pseudo code for searching on B+ANN indexing.

\begin{algorithm}[hbt!]
\caption{B+ANN Search}\label{alg:search}
\small
\begin{algorithmic}[1]
\Procedure{Search}{$\mathbf{q}$, $k$, $\beta$, $J$, $d_{\mathrm{edge}}$}
    \State $outputs \gets [\,],\; visited \gets [\,]$
    \State $Q \gets \mathrm{Queue}(\text{root}),\; PQ \gets \mathrm{PriorityQueue}()$
    \While{$!Q.\mathrm{empty}()$}
        \For{$node \in Q.\mathrm{pop\_all}()$}
            \State $dist \gets \mathrm{dist\_1d\_2d}(node.vectors, \mathbf{q})$
            \State $idx \gets \mathrm{argsort}(dist, k)$
            \If{$node$ is internal}
                \For{$i \in 1..\beta$} 
                    \State $PQ.\mathrm{push}(node.child(idx[i]))$
                \EndFor
            \Else
                \For{$i \in idx$}
                    \State $v \gets node.vectors[i]$
                    \If{$visited[v.id] = 0$} 
                        \State $outputs.\mathrm{add}(v),\; visited[v.id]\gets1$
                    \EndIf
                \EndFor
                \If{$d_{\mathrm{edge}} > 0$}
                    \State $\mathrm{GreedySearch}(\mathbf{q}, outputs, k, visited)$
                \EndIf
            \EndIf
        \EndFor
        \State $\mathrm{SyncLRU}(Q)$
        \State $Q \gets PQ.\mathrm{pop\_top}(\beta)$
    \EndWhile
    \State \Return $outputs$
\EndProcedure
\end{algorithmic}
\end{algorithm}

\subsection{Building and Traversing the Skip Connections}
\label{sec:skip-conn}

Tree indexing is the first step in efficiently managing and indexing large volumes of data. To make precise searches, we need finer granularity. As we mentioned earlier, the data can be clustered in the local hubs with thousands of vectors. Therefore, the partitioning at the dense level creates tight regions with a high population of vectors lying within the edges, and querying the nearest object would end up in the wrong region, resulting in suboptimal retrieval, which is known as the \textit{edge problem}. For finer granularity, we introduce \textit{skip-edges} to traverse between neighboring regions and circumvent the edge problem. As shown at Phase-2 of the Figure \ref{fig:main_fig}, we are connecting a vector within its leaf node, \textit{inter-connection}, and also a neighbor node, \textit{intra-connection}. The inter-connections represent hops between nodes, whereas the intra-connections enable local searches among a node’s nearest vectors whenever a hop is performed. The introduction of skip-edge connections transforms the tree into a graph, yielding a hybrid structure that leverages the complementary advantages of both representations. Next, we describe how skip-connections are specified, how they are created and how we use them in the greedy search. 

\textbf{Building Skip-Edges:} The skip-edge connections are governed by two parameters: $d_{edge}$, which specifies the connection degree determining the number of links established for each vector at the leaf nodes, and $s_{\mathrm{leaf}}$, which defines the size of the leaf set explored during the connection construction process. Constructing the skip-edge connections requires identifying the nearest leaf nodes for each node. Thereby, we first compute the top-
$s_{\mathrm{leaf}}$ nearest leaf nodes for every node based on pairwise distances between their centroids. Let $N_{\mathrm{leaf}}$ be the total number of leaves, then it requires calculating $N_{\mathrm{leaf}}^2$ pairwise distances of each leaf-node. The number of leaf nodes is determined by the capacity of the leaf nodes, and the expected number of leaves is $N_{\mathrm{leaf}}\approx \lceil \nicefrac{N}{\kappa_{\mathrm{leaf}}} \rceil$, making the total complexity $O(\nicefrac{N^2}{\kappa_{\mathrm{leaf}}^2})$. If we select $\kappa_{\mathrm{leaf}} = \sqrt{\nicefrac{N}{\log N}}$ then the total complexity reduces to $O(N\log(N))$, for $N=10^6$ setting $\kappa_{\mathrm{leaf}} = 270$ nodes would be enough to reach this complexity. The computation is accelerated through multi-threading and SGEMM operations, reducing execution time to a few seconds owing to its fully parallelizable nature.

After identifying the top-$s_{\mathrm{leaf}}$ closest leaf nodes, intra-connections are established by selecting, for each vector in a leaf node, the top-$d_{\mathrm{degree}}$ closest vectors among those contained in the top-$s_{\mathrm{leaf}}$ leaves. Each vector in a leaf node connects to its top-$d_{\mathrm{degree}}$ closest vectors among the nearest leaves. Compared to HNSW, these connections are shorter decreasing the diameter of the graph and enabling it to reach the closest vector more quickly. Since we calculate distances to each vector among top-$s_{\mathrm{leaf}}$ closest leaf nodes, the complexity is $O(\kappa_{\mathrm{leaf}}s_{\mathrm{leaf}}N)$. When we assume $\kappa_{\mathrm{leaf}}= \sqrt{\nicefrac{N}{\log N}}$ and if $s_{\mathrm{leaf}} \leq \sqrt{N\log(N)}$ then overall complexity is strictly less than $O(N^2)$ for building the connections. For $N=10^6$ setting $s_{\mathrm{leaf}} = 2450$, as we show in our experiments, usually 512 value is enough to reach above 90\% recall value.

\begin{algorithm}[hbt!]
\caption{Traversing the Skip-edges}\label{alg:greedy_search}
\small
\begin{algorithmic}[1]
\Procedure{GreedySearch}{$\mathbf{q}$, $outputs$, $k$, $visited$}
    \State $L\gets outputs,\; L' \gets outputs$
    \For{$v \in outputs$, where $visited[v.id] \leq 1$}
        \For{$u \in v.edges$ where $visited[u.id] < 1$}
            \State $L.\mathrm{add}(u)$, $u.id \gets 1$
        \EndFor
         \State $v.id \gets 2$ \Comment{node $v$ is all discovered}
    \EndFor
    \State $L\gets \mathrm{get\_nearest\_k}(\mathbf{q}, L, k)$ \Comment{Retain only k closest points to $\mathbf{q}$}
    \If{$L'\; != L$}
        \State GreedySearch($\mathbf{q}$, $L$, $k$, $visited$)
    \EndIf
    \State \Return $L'$
\EndProcedure
\end{algorithmic}
\end{algorithm}

\textbf{Greedy Search:} The traversing algorithm, \textit{greedy-search}, is shown in Algorithm \ref{alg:greedy_search}. We adapt the greedy-search strategy with an important change: We initialize the set of visited vectors, $L$, with the $outputs$ of the tree search, as shown at 20\textsuperscript{th} line in Algorithm \ref{alg:search} and at the 2\textsuperscript{nd} line in Algorithm \ref{alg:greedy_search}. This allows multiple search seeds to be initialized concurrently during multi-threaded execution, and converge the final set more rapidly by traversing at different branches in parallel, as shown in Figure \ref{fig:graph_traversal}. The visited array is an atomic state variable that can take values 0, 1, or 2, each representing a distinct visitation state: a value of 0 indicates that the vector has not been visited, 1 denotes that it has been visited but its edges have not yet been explored, and 2 signifies that all its edges have been fully discovered. For each vector in the $outputs$, we discover its edges and put them into $L$. Then we retain the top-$k$ nearest vectors in $L$. If there is a new vector added to the initial set $L$, then we recursively call the greedy search again. If there are no changes in the output vectors, then we stop traversing and return the final $L$.

\textbf{Usage of Skip-Edges:} During the indexing of a dataset $\mathcal{D}$, skip-edge connections are not constructed by default; instead, the user may optionally enable their creation through a configuration parameter. The created graph structure is maintained in memory, yet it can also be serialized to the disk for persistence. Compared to the tree, skip-edges incur a higher computational cost to build, we expose this option as a user-defined argument to enable finer-grained search when required. In the following section, we utilized this feature to first index extreme datasets ($|\mathcal{D}|\gg \text{billions}$) by tree indexing and creating view representations with skip-edge connections. 

\subsection{View Creation and B+ANN Usage}

As we defined in Section \ref{sec:problem}, for a query $\mathbf{q}_t$ at time $t$ we extract the view from the B+ANN Tree by $V(\mathbf{q}_t, \mathcal{T}, k) = \mathcal{T}'$ and obtain the graph created $G' = (\mathcal{L}(\mathcal{T}'), \mathcal{E})$ by connecting the leaves of sub-tree $\mathcal{L}(\mathcal{T}')$ with the skip-edge connections, where $\mathcal{E}$ represents the edges between pairs of vectors. Then, the subsequent queries $\mathbf{q}_{t+1}, \mathbf{q}_{t+2}, \dots$ are served this representation using B+ANN search (Algorithm \ref{alg:search}). If the retrieved $k$-ANN vectors of any of the subsequent queries do not exist in the true $k$-NNs, i.e., recall hits to zero, we extract another view. We call the duration between two succeeding views as the \textit{view survival time}. As shown in Figure \ref{fig:seq_motivation_b}, when the queries are temporarily correlated, the survival time of a view can last up to 1000 queries. Moreover, the duration can be increased by setting the $k$ parameter of the view function to a high value (e.g., 1000) to populate a view that lasts longer. 

The extraction of a view can be applied to a B+ANN tree $\mathcal{T}$ residing either in memory or on disk. Since the operation relies on efficient tree-based search, retrieving $k$-ANNs ($k<1000$) is extremely fast, and constructing a view with skip-edge connections for $N<1000$ nodes typically completes within seconds. The value 1000 here denotes an arbitrarily large constant. Most importantly, when $\mathcal{T}$ resides in the disk, it can asynchronously be updated by inserting new vectors, while the extracted view $\mathcal{T}'$ responds to the incoming queries. Separating the computation to disk and memory allows us to store a clean, lightweight data structure on disk with locality-improved storage that allows fewer cache misses, and we move the fine-grained computation to the memory. Therefore, we can store a gigantic amount of data in a disk without any spurious and complex components, such as edges in a graph indexing.

Another important aspect of view creation is its semantic representation. It represents a collection of semantically related information of a query vector, and when this query is related to a particular topic, the view encapsulates the most relevant information of that topic. This opens the door for various applications. In terms of access control, a user can have access to a particular view of data that represents a subset of the information through a specific query. Moreover, it is possible to extract multiple views from a dataset and use the information provided by them as an expertise in different topics. Therefore, the area confined by the view makes it possible to search for the ranges of particular interest, which is useful for especially dissimilar queries. Once a view is extracted for a query, we define a context, and we can find the most dissimilar queries within that context. For example, if we are finding the most dissimilar "cereal" of a given cereal item, in ordinary VDBs, we would retrieve highly unrelated items. On the other hand, by the view creation of the cereal item, we determined the border of our search with a context and can retrieve the most dissimilar "cereal" for the given item. Next, we evaluate the performance of B+ANN.

\section{Evaluation}
\label{sec:eval}

\subsection{Datasets and Experimental Setup}
In our experiments, we have used two benchmark datasets: The first is
\texttt{Glove-1.2M} containing 1.2 million 100- and 200-dimensional
word embeddings introduced by \cite{pennington2014glove}. We indexed
the vectors in this dataset based on the cosine distance and calculate
the recall value of the same test split of \cite{ann-bench:2018}. To
observe the scalability and the performance of Euclidian Distance, we
utilized \texttt{SIFT-1B} dataset from \cite{Amsaleg2010Texmex}
containing 1 billion 128-dimensions of image descriptors. Furthermore,
to run low scale experiments such as dissimilarity search, we used the
Spotify dataset \cite{annoy} containing 243k vectors with 256
dimensions.  

We compare two Instruction Set Architectures, Arm64 and x86. For
Arm64, we used an Apple M4 Max system with 16 Performance cores and 32
GB of unified memory, running without virtualization. For the x86
architecture, we employed a workstation virtualization with Intel Xeon
Platinum CPUs featuring 128 cores and 64 GB of memory, where we also
measured the utilization of L1 and L2 caches.  

For implementation, we used C++17 for our codebase and the CBLAS
library to perform SIMD operations to calculate distance metrics. As
an evaluation metric, we followed standard benchmark set-up from
\cite{ann-bench:2018} and observed speed-recall trade-off by measuring
10-Recall ($\%$) and Query Per Second (QPS) ($1/s$) performance.  

\subsection{In-Memory Performance of B+ANN}
\begin{figure}[hbt!]
    \centering
    \includegraphics[width=0.95\linewidth]{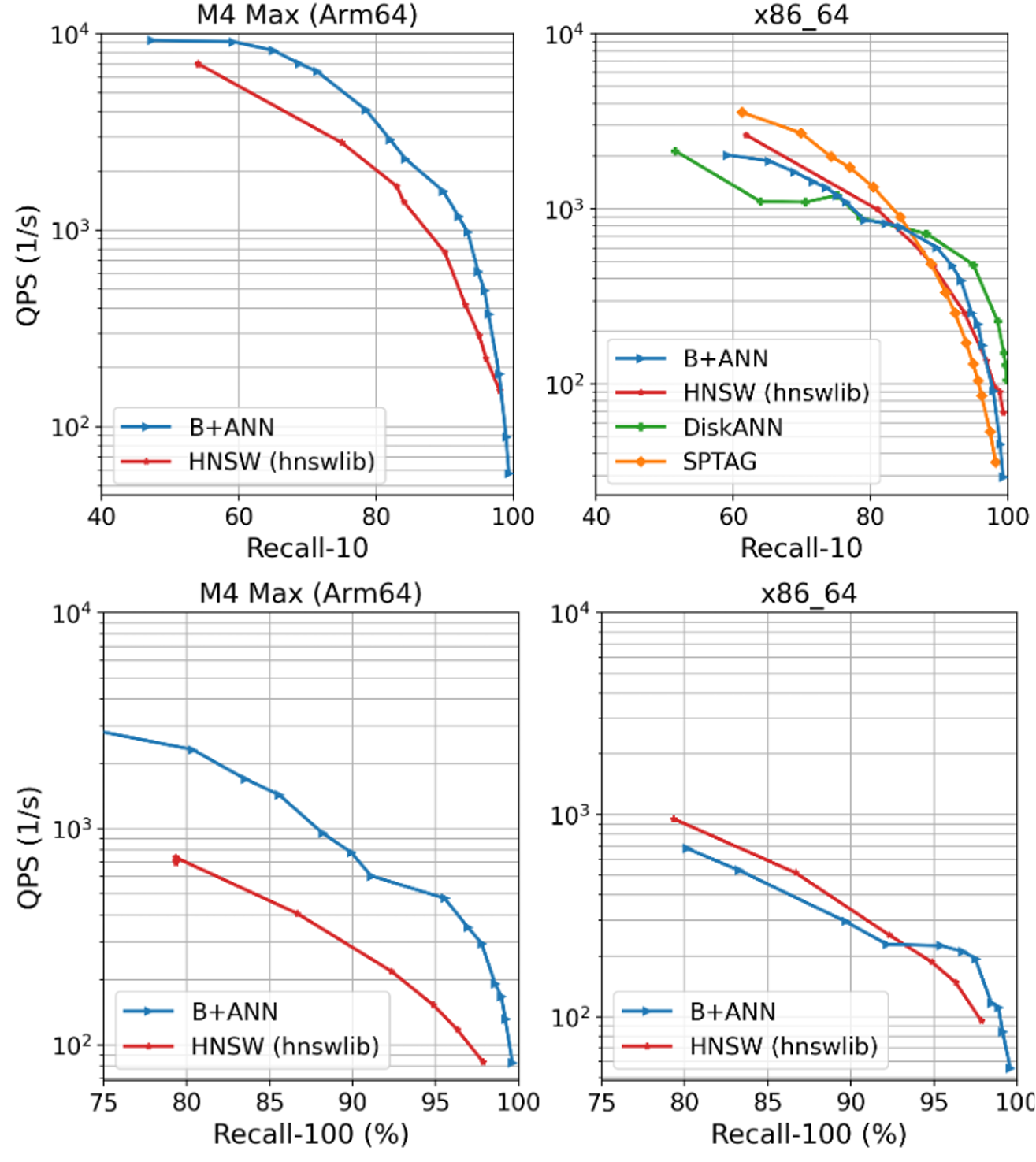}
    \caption{In-memory performance of ANN algorithms and B+ANN. We show QPS vs Recall-10 and Recall-100 curves of benchmark algorithms and B+ANN for Arm64 and x86 architectures.}
    \label{fig:main_results}
\end{figure}

Based on this setup, we measured the performance of B+ANN and multiple
benchmark algorithms. We selected HNSW as the main graph-based
algorithm to observe the performance change by our design based on our
motivation in section \ref{sec:preliminaries}. We enriched the
comparison with the two SOTA disk-based ANN algorithms, DiskANN
\cite{jayaram2019diskann}, and SPTAG \cite{ChenW18}. While DiskANN is
built on a sparse neighborhood graph, SPTAG is also a hybrid approach
that utilizes space partition and multiple Relative Neighbourhood
Graphs. Our HNSW experiments were conducted with \texttt{hnswlib}
\cite{malkov2018hnswlib}, a C++ implementation. While hnswlib and
B+ANN are architecture-agnostic, DiskANN and SPTAG are tied to
architecture-specific features, and they require containerization to
be implemented in other architectures. All the baseline methods can
perform SIMD accelerations and utilize multi-threading.  

In this section, we compared the in-memory performance of the used
SOTA algorithms with B+ANN. Therefore, we build the Hierarchically
Clustered B+Tree in memory and create the skip-edge connections
between nodes. We make an analysis of the disk performance in section
\ref{sec:disk_perf}. Following \cite{jayaram2019diskann} parameter
selection, for all the algorithms, during construction, we set the
edge degree to $M=128$, and the effective search parameter to
${ef}_{C}=512$. For the architecture-specific parameters, e.g., Vamana
indices $R$, $L$, and penalty constant $\alpha$, we use the default
parameters listed on their corresponding repositories. Accordingly, we
set the parameters of B+ANN as follows: ${\kappa}_\mathrm{leaf}=2048$,
${\kappa}_\mathrm{inner}=1024$, $d_{edge} = 128$, and $s_{leaf}= 512$,
and built the tree and created the skip-connections all in
memory. Figure \ref{fig:graph_built} shows the graph built time for
all the algorithms in different architectures. We report the time it
takes to build the tree and skip-edges. The result shows that B+ANN
has the fastest build time in arm64 and s390x architectures, while
having the third fastest build time in x86. SPTAG, on the other hand,
took more than 3 times the build time of B+ANN. 

After the indexing, we perform search multiple times by changing the
search width parameter, e.g. $ef$ of algorithms ranging from $5$ to
$1200$ and measure the recall values. The corresponding parameter in
B+ANN is the branching factor, $\beta$, which ranges from $5$ to
$100$. Lastly, all algorithms were executed using the same number of
threads, $threads=10$. We also performed experiments keeping the
thread count to 1, however, B+ANN outperforms the other algorithms
with a larger gap when $threads>1$ showing the importance of cache and
memory locality.  

Figure \ref{fig:main_results} shows the speed-recall curves of the
algorithms obtained using the described parameters and experimental
setup. Compared to HNSW, B+ANN achieves a $10\times$ speedup at
10-Recall and a $50\times$ speedup at 100-Recall in arm64 architecture
and $1.1\times$ to $2 \times$ speed up in x86, while showing the same
recall high value. Relative to DiskANN and SPTAG, B+ANN shows
comparable performance yet these algorithms require CPU specific
speed-ups e.g. AVX512. Next, we show that B+ANN is much more
advantageous for disk implementation. 

\begin{table}[t]
  \begin{adjustbox}{width=0.49\textwidth, center}
    \centering
    \small
    \begin{tabular}{l M{1.5cm} M{1.5cm} M{1.5cm} M{1.5cm}}
    \hline
    \multicolumn{5}{c}{Architecture Type \textbf{x86\_64}}\\
    \hline
    Method & IPC ($1/s$) & L1 Load Miss $(\%) \downarrow$ & Branch Miss $(\%) \downarrow$ & QPS $(1/s) \uparrow$ \\
    \hline
    HNSW & $0.68$ & 19.76 & 3.19 & 186.64 \\ 
    B+ANN & 0.29 & 15.96 & 3.21 & 225.19 \\
    \hline
    Rel. Gain & - & \textcolor{darkgreen}{$-19.23\%$} & {$+0.62 \%$} & \textcolor{darkgreen}{$+20.65 \%$} \\
    \hline
    \end{tabular}
    \end{adjustbox}
    
    \caption{We measure multiple performance statistics in x86 and s390x for HNSW and B+ANN at the $95\%$ 10-Recall value. 
    }
    \label{table:gsm8k}
    \vspace{-12pt}
\end{table}

\subsection{On-Disk Performance and Memory Usage}
\label{sec:disk_perf}

\begin{figure}[htbp]
    \centering
    \begin{subfigure}[b]{0.45\textwidth}
        \centering
        \includegraphics[width=0.75\textwidth]{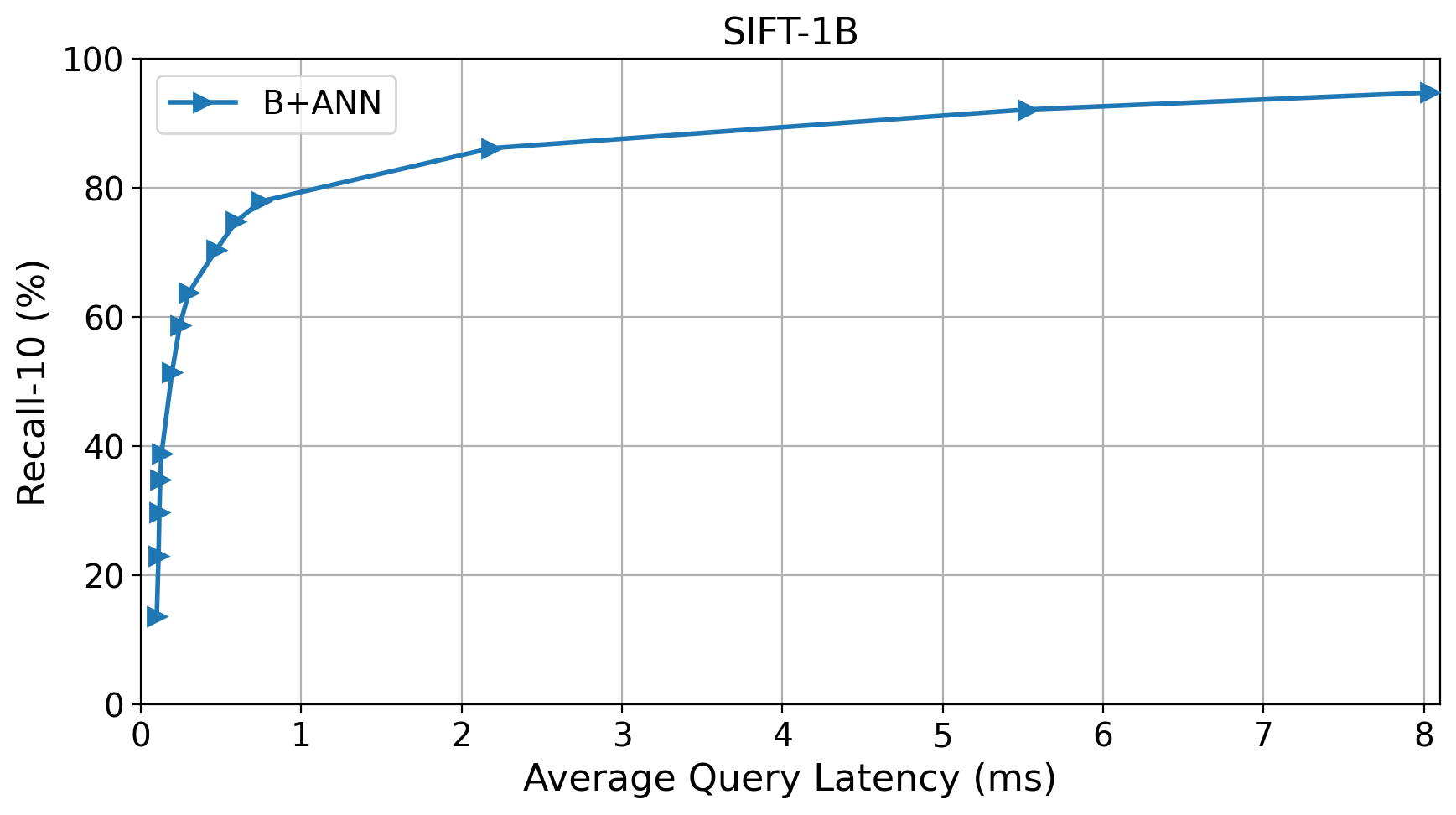}
        \caption{The Recall-10 vs latency of on SIFT-1B dataset. }
        \label{fig:sift}
    \end{subfigure}
    \hfill
    \begin{subfigure}[b]{0.49\textwidth}
        \centering
        \includegraphics[width=\textwidth]{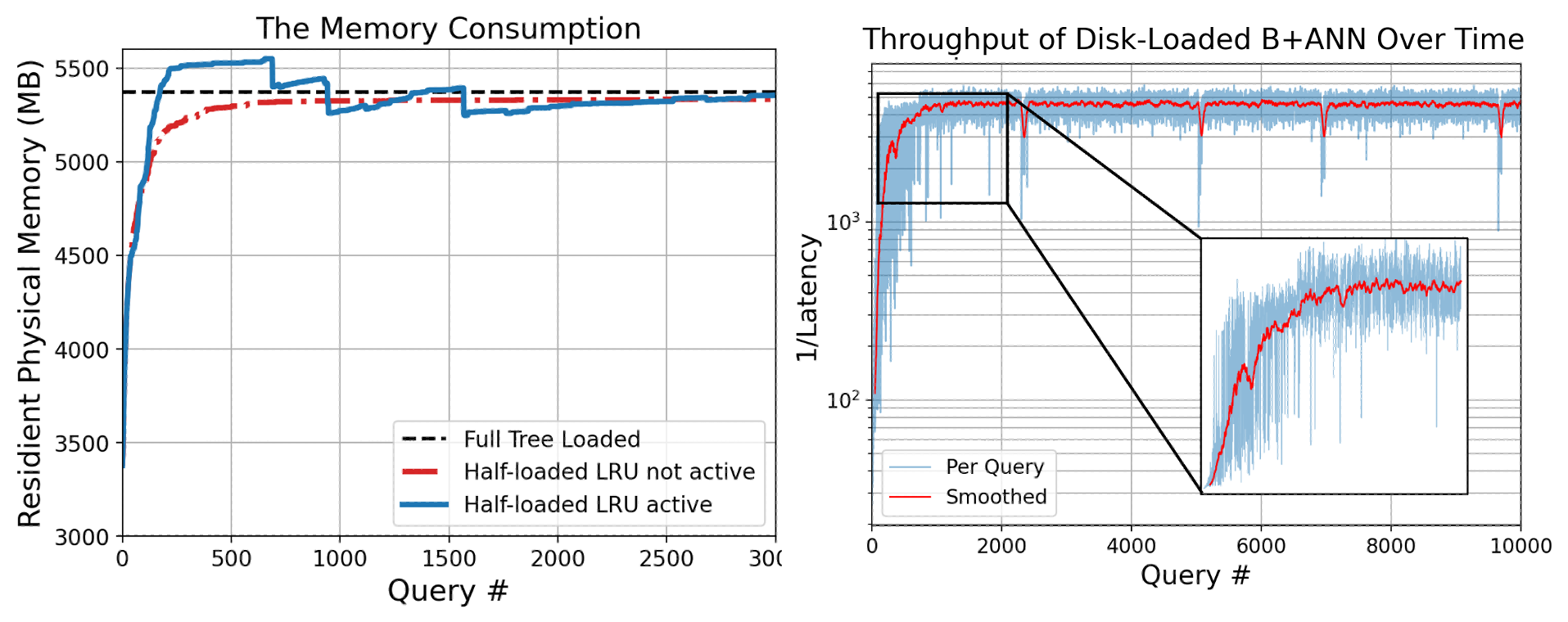}
        \caption{The left shows how LRU balances the memory consumption when a half-loaded (half of the nodes are in memory and the other half is on disk) tree is used. The right shows how B+ANN balances the throughput as it performs search by loading nodes from disk to memory (it starts with zero-loaded tree).}
        \label{fig:mem_tpt}
    \end{subfigure}
    \caption{B+ANN Performance for the SIFT-1B dataset}
    \label{fig:memfigs}
\end{figure}

For the on-disk experiments, we utilized the \texttt{SIFT-1B} dataset
and built the B+ANN tree on the disk, which took 2 hours with $150$GB
maximum memory usage. This is $24\times$ faster compared to DiskANN
built time and $350$GB less memory usage\cite{jayaram2019diskann}. The
reason is that B+ANN tree building time has the complexity of $O(n\log
n)$. The search performance is shown in Figure \ref{fig:sift}. The
plot shows that B+ANN can reach 99.8\% Recall-10 in less than 9ms and
80.0\% recall in less than 1ms. The results show that B+ANN can
respond to queries within milliseconds with high precision, even for
huge datasets without requiring full in-memory indexing, thereby
drastically reducing hardware cost and memory footprint. This
capability is critical in applications such as large-scale
recommendation engines, semantic search systems, and knowledge
retrieval platforms, where responsiveness and accuracy jointly
determine user satisfaction and system scalability. 

Figure \ref{fig:mem_tpt} shows our second analysis of disk-based
search in B+ANN. During the search, B+ANN uses hybrid memory usage by
loading the leaf-nodes from disk to memory. Initially, all the nodes
are located on disk, and the memory capping is active. As we perform
the vector search by traversing the levels of the tree, B+ANN loads
the nodes to memory through LRU. For example, the Figure
\ref{fig:mem_tpt} shows the memory consumption and throughput as we
perform searches on a B+ANN index tree with half of the nodes located
on disk and the other half located in memory. We make two
observations: (i) The LRU queue balances the memory usage when the
memory exceeds the threshold, and it asynchronously dumps the least
recently used nodes into the disk. A natural question is, why can't we
let this process be handled by the OS and its swapping mechanism,
especially in macOS, since everything operates on the same memory
fabric? The reason is that when the OS swaps inactive memory, the
paging and addressing are handled by the OS with respect to its own
grouping methods; however, B+ANN has a structured storage system that
already maximizes the locality. It keeps semantically close vectors
(and nodes) also physically close on disk, thereby increasing locality
and faster I/O operations. (ii) We observe that as the searching continues, 
the throughput exhibits asymptotic
growth. Initially, the number of I/O operations is high due to the
tree being fully located inside the disk. As we index queries, we load
nodes from disk to memory, and the probability of nodes we fetch from
the memory (LRU queue) increases; consequently, the latency
decreases. The results exemplify the robust convergence property of
the B+ANN, attaining a steady operating point, ensuring long-term
predictability and stability.

\subsection{Average Hops Comparison}
\begin{figure}[hbt!]
    \centering
    \includegraphics[width=0.9\linewidth]{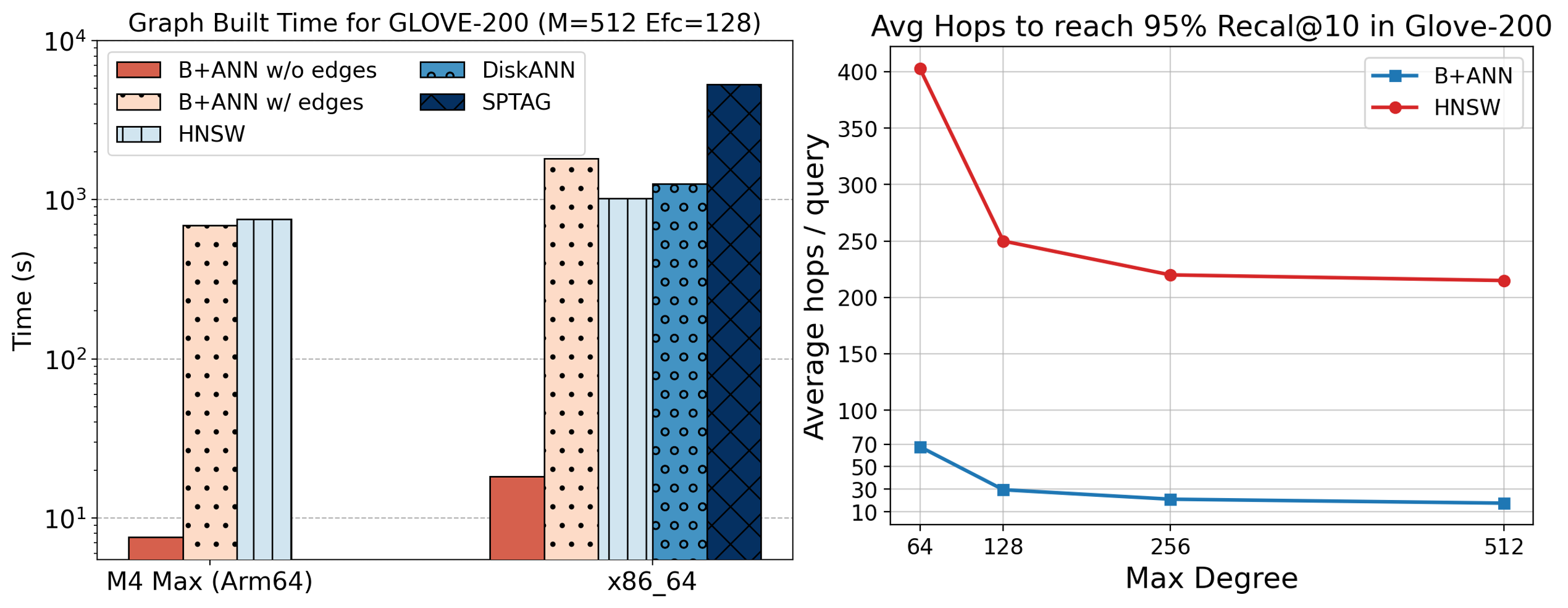}
    \caption{Figure on left shows the graph build time comparison of each algorithm for different architectures. The plot on right shows the average-hop comparison between B+ANN and HNSW (lower is better).}
    \label{fig:graph_built}
\end{figure}

The plot on the left in Figure \ref{fig:graph_built} shows the
graph-building time of baseline indexing algorithms and B+ANN. We
report the total time taken for B+ANN to build the Hierarchical
Clustering B+Tree and the skip-edge connections. Also, we calculated
that it takes $84.3\%$ of total indexing time to build skip-edge
connections. The difference is because the complexity of building the
tree is $O(n\log(n))$ lower than creating the skip-edge connections,
$O(n_{leaf}.n.d_{edge})$. However, the complexity of building the
skip-edge connections can be reduced by adding the vectors to the
leaf-nodes in an ordered way, e.g., based on their distance to their
centroid. Connecting vectors to the closest vector in the neighboring
node, therefore, would require less computation. Based on the scope of
our paper, we leave it as future work. 

To test the diameter of skip-edge connections, we compare the number
of visited nodes in an HNSW graph with the number of visited nodes in
B+ANN with skip-edge connections. The Figure \ref{fig:graph_built}
shows the average hop comparison between HNSW and B+ANN to reach
Recall-10 using set of degrees $d_{edge} = \{64, 128, 256, 512\}$
building B+ANN and corresponding $M$ parameter for HNSW. Then we
increased the effective search parameter, $ef$, for HNSW and branching
factor, $\beta$, for B+ANN until we observed an average $95\%$ Recall
value. The plot shows that B+ANN uses $15\times$ fewer hops than
HNSW. Based on this observation, B+ANN creates shorter connections,
thereby decreasing the diameter and enabling it to reach the closest
vector more quickly. Fewer hops also decreases the number of I/O,
which is critical during hybrid memory search and when memory capping
is active.

\subsection{Exploiting Temporal Correlation by B+ANN}

\begin{figure}[hbt!]
    \centering
    \includegraphics[width=0.65\linewidth]{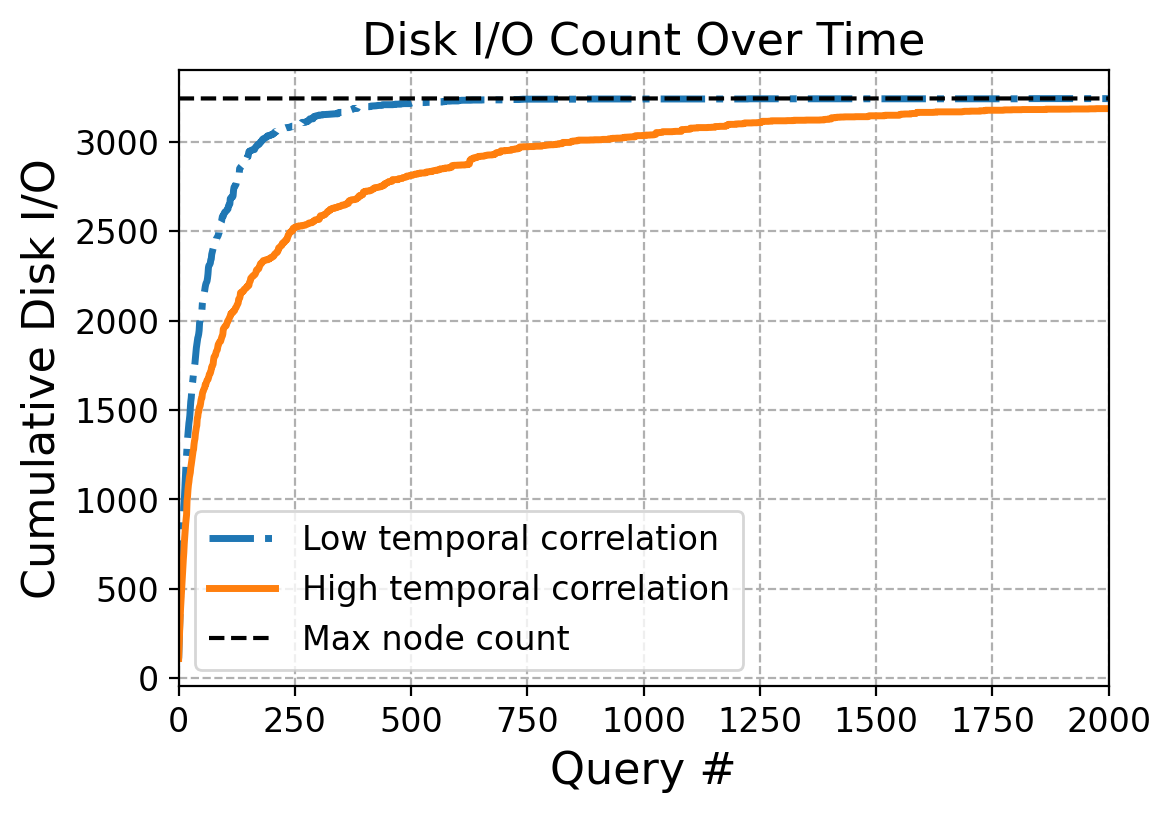}
    \caption{Plot shows the cumulative number of Disk I/O operations performed when there is a high temporal and low temporal correlation in SIFT-1B dataset.}
    \label{fig:temporal_corr_high}
\end{figure}

One of the biggest advantages of B+ANN appears when the input queries
are temporally correlated. To simulate this, we have sorted the
incoming queries based on their distances to each other. For a query
at time $t$ denoted by $q_t$, the next query is within proximity yet
it has not appeared before, represented as $q_{t+1}\in R(q_t)$ and $
R\subset S\setminus Q_{\mathrm{prev}}$, where $R(q_{t})$ is the
vectors at proximity of $q_t$ computed on the unused queries and
$Q_{\mathrm{prev}}=\{q_1,\dots,q_{t-1}\}$ is the set of previous
queries before time $t$. The Figure \ref{fig:temporal_corr_high} shows
the total Disk I/O count as the number of queries with high temporal
correlation compared to random order. In randomly selected queries,
the graph is discovered more quickly by visiting the unvisited nodes
more frequently. In contrast, high temporal correlation shows a slower
increase in Disk I/O count compared to the random selection. The
answer to the next query can be retrieved from the node that was
already put in memory in the earlier responses. The critical impact of
this property is also observed in Figure \ref{fig:seq_motivation_a},
the survival time of the view can reach up to 1000 queries due to
responding from the created view in memory, and the incoming query
will be in the view. Therefore, if temporal correlation exists, B+ANN
will eventually delay the new view creation and reduce the average
response time. 

\subsection{Balanced Growth and Insertion}

\begin{figure}[htbp]
    \centering
    \includegraphics[width=0.46\textwidth]{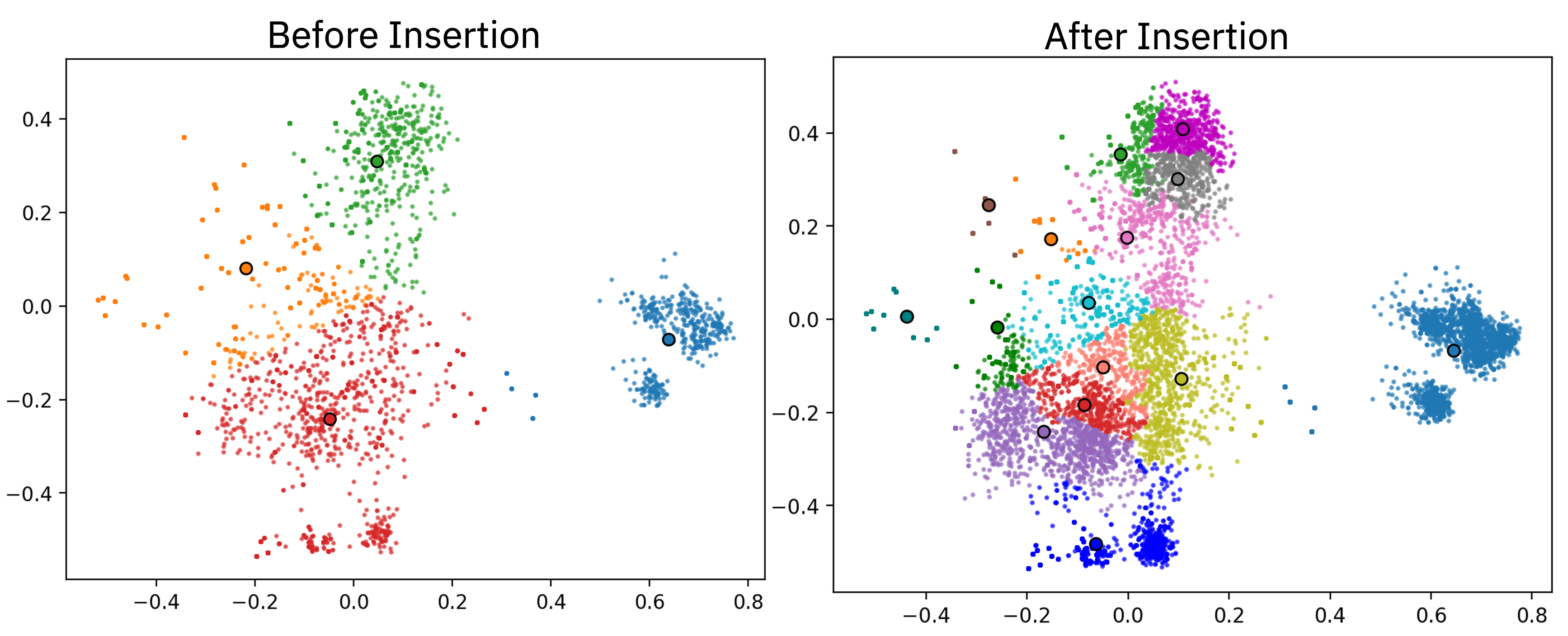}  
    \caption{Visualization of the clustering after insertion: initial cluster mapping is updated as new vectors are added (e.g., the
before-insertion green cluster gets split into 4 different clusters).}
    \label{fig:observation}
\end{figure}

For an indexing algorithm, being flexible and updatable is essential,
especially in systems that terabytes of fresh data continuously
coming. One the reasons of selecting B+ Tree architecture is this
flexibility and being able to grow in width without performance
depreciation. In B+ANN, the incoming vector is inserted to one of the
leaf nodes and the leaf is split if it reaches the capacity. This
split operation performed using K-Means, and the vectors inside that
leaf node are divided into two clusters creating two new centroids
that needs to be added into inner nodes.  The number of vectors that
needs to be inserted for a node to split increases exponentially as we
go from lower levels to higher levels. Hence, for an input vector, the
update mechanism locally adjusts branches without drastic update on
overall tree index structure. Figure \ref{fig:observation} illustrates
the growth in the B+ANN. We performed PCA on the vectors that are
indexed to visualize in two-dimensional space and observe the clusters that are created
and assigned. On left, only 15\% of the data is indexed which has 4
leaf nodes. On right, rest of the data is inserted and the growth on
the clusters are shown. Observe that the clusters are separated
w.r.t. their previous location, e.g., the green cluster on left is
subdivided into 4 clusters respecting the distances. Secondly, observe
the blue cluster on the right side of Before-Insertion and
After-Insertion plots. Vectors in the vicinity of the blue cluster are
assigned to that cluster, and the corresponding node is updated
without a split operation since its capacity is not reached.

\subsection{Impact of Batch Size on Performance}

\begin{figure}
    \centering
    \includegraphics[width=0.65\linewidth]{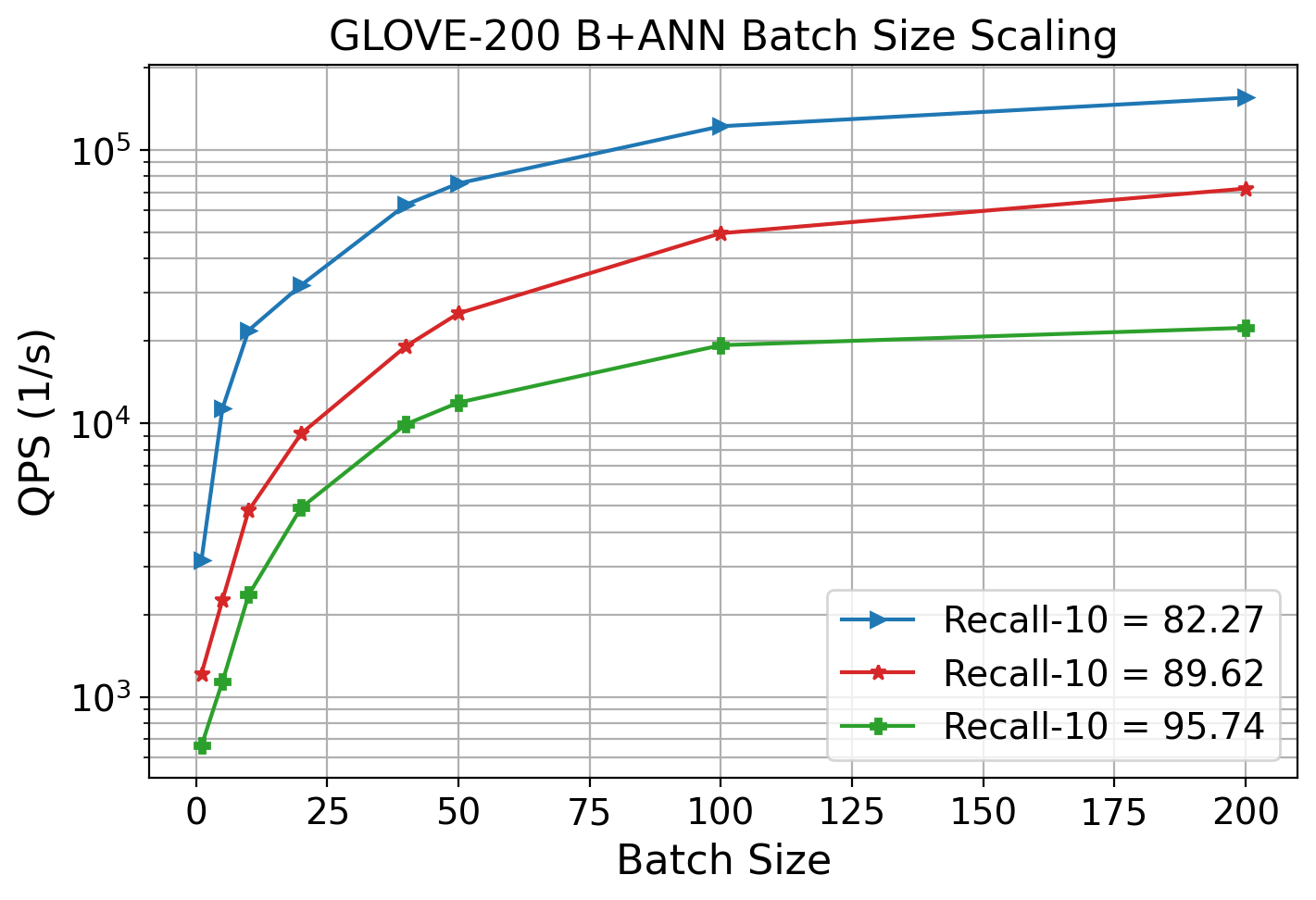}
    \caption{Scalability of query throughput (QPS) with respect to
      batch size, in GLOVE-200. Larger batch sizes generally improve
      throughput until saturation is reached.} 
    \label{fig:batch}
\end{figure}

\begin{figure}
    \centering
    \includegraphics[width=\linewidth]{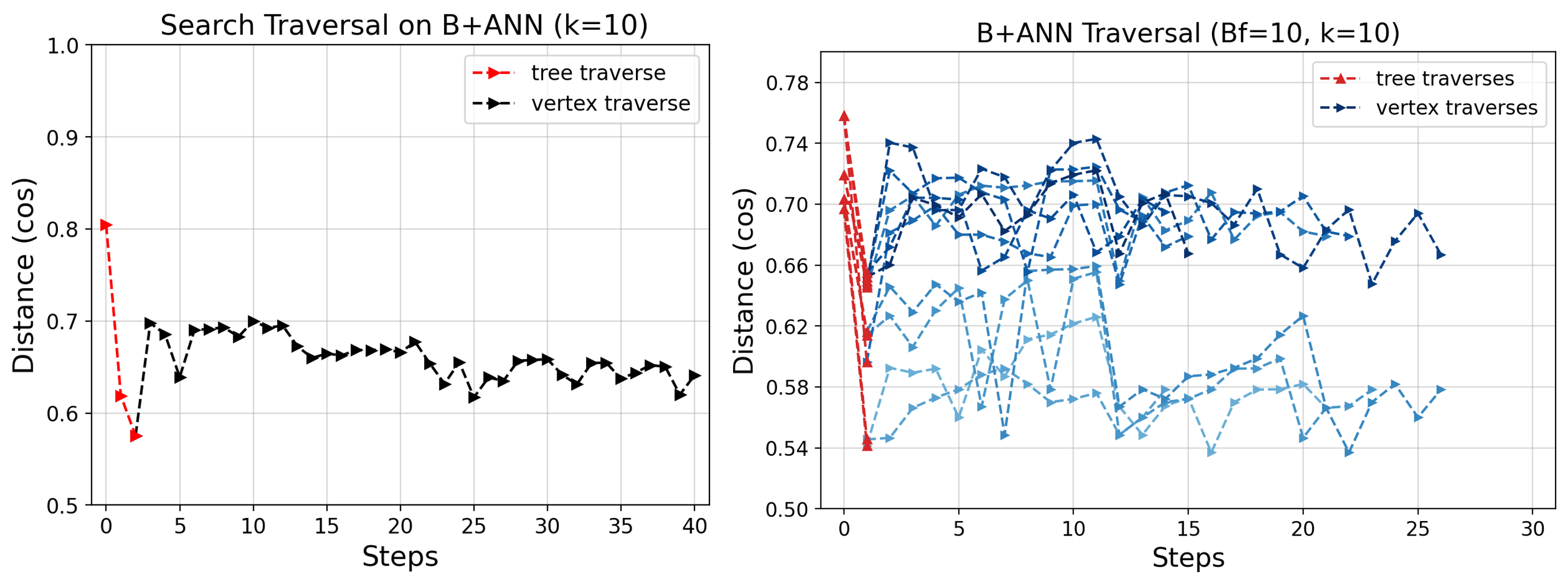}
    \caption{We show the B+ANN traversal for a query and a batched query}
    \label{fig:graph_traversal}
\end{figure}

The Figure \ref{fig:batch} plots the relation between QPS and batch
size and testing the scalability of B+ANN as the workload gradually
increased. The colored lines indicate configurations with a constant
Recall-10 value, while the batch size is varied and all other
parameters are held fixed. Here, we calculated
$\mathrm{QPS}=\frac{N_{\mathrm{queries}}}{T_{\mathrm{total}}}$. When
queries are batched, vector–matrix operations are replaced by
matrix–matrix operations during tree traversal, which is shown in
Figure \ref{fig:graph_traversal}. Consequently, the scalability of
tree traversal depends on CBLAS SGEMM efficiency, which exhibits exponential
growth. In contrast, during skip-edge traversal, each query vector in
the batch is processed by an individual thread, limiting scalability
to the available number of threads. Therefore, all the lines in Figure
\ref{fig:batch} shows asymptotic growth with rapid increase initially
and attain a steady QPS thereafter. Moreover, we observe up to
$\times100$ increase in QPS as we increase the batch size. 

\subsection{Evaluating Dissimilarity Search}

\begin{figure}[hbt!]
    \centering
    \includegraphics[width=0.65\linewidth]{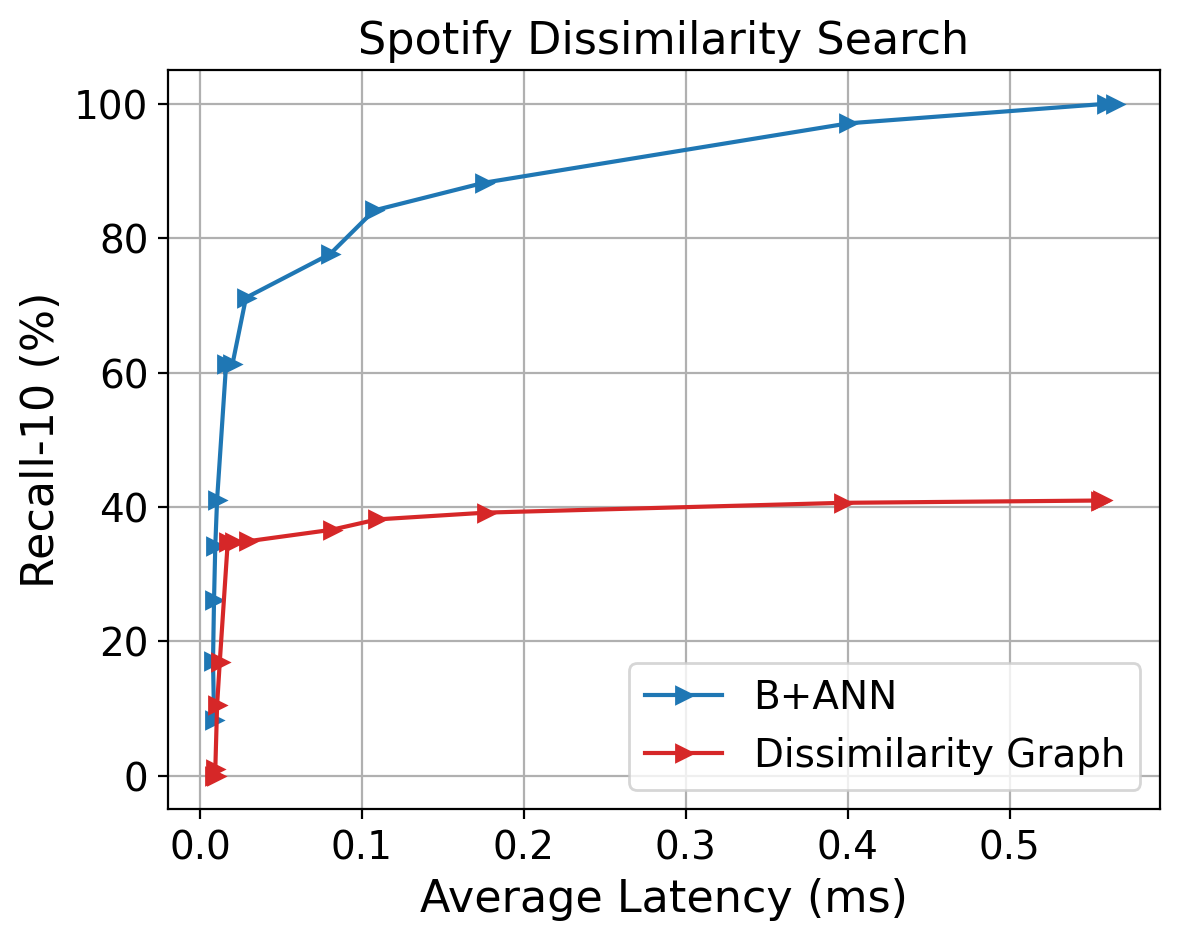}
    \caption{We evaluate the dissimilarity performance of B+ANN
      relative to a baseline dissimilarity graph (higher is better).}
    \label{fig:dissimilarity}
\end{figure}

Unlike the traditional similarity queries, dissimilarity queries are
used to indentify semantically different entities. For example, in Db2
SQL Data Insights, dissimilarity queries are often used to extract
outliers~\cite{sqldi}. 

To measure the performance of the dissimilarity search functionality
of B+ANN, we reversed the problem by a labeled dataset where each
vector has a label set that contains the IDs of the top-k farthest
vectors. Using this setup, we query the B+ANN using the dissimilarity
flag and measure the Recall-10. Then, we developed a new baseline
algorithm, as dissimilarity-based vector search has been sparsely
studied in the literature, and current HNSW-type graph algorithms
cannot support dissimilarity queries. We created a Relative
Neighborhood-graph (RNG) based on the distance between vectors, and we
connect two vectors if they are farther from each other than any other
vectors. The performance of this baseline and B+ANN is shown in Figure
\ref{fig:dissimilarity}. B+ANN can reach $100\%$ Recall within less
than 0.6ms, while dissimilarity-RNG can reach up to 40\%. We observe
that dissimilarity-RNG oscillates between the same graph nodes,
therefore converges to a local solution. 


\section{Conclusion}
\label{sec:conclusion}
This paper presents B+ANN, a novel disk-based ANN algorithm, the
improves functionality, result accuracy, and performance over HNSW,
currently the most used ANN algorithm in Vector Databases. Our
extensive performance evaluation using a variety of large vector
datasets demonstrates that B+ANN improves in-memory and disk access
patterns leading to improved spatial and temporal locality, leading to
improved computation acceleration opportunities. The B+ANN tree
approach also enables efficient execution of dissimilarity queries,
which are not supported by HSNW-based indices. As a future work, we
want to further explore the notion of semantic views and its impact on
the index architecture. We also want to explore opportunities for
memory reduction using low-precision floating point representations.

\bibliographystyle{ACM-Reference-Format}
\bibliography{main, ai-related}


\begin{thebibliography}{60}


\ifx \showCODEN    \undefined \def \showCODEN     #1{\unskip}     \fi
\ifx \showDOI      \undefined \def \showDOI       #1{#1}\fi
\ifx \showISBNx    \undefined \def \showISBNx     #1{\unskip}     \fi
\ifx \showISBNxiii \undefined \def \showISBNxiii  #1{\unskip}     \fi
\ifx \showISSN     \undefined \def \showISSN      #1{\unskip}     \fi
\ifx \showLCCN     \undefined \def \showLCCN      #1{\unskip}     \fi
\ifx \shownote     \undefined \def \shownote      #1{#1}          \fi
\ifx \showarticletitle \undefined \def \showarticletitle #1{#1}   \fi
\ifx \showURL      \undefined \def \showURL       {\relax}        \fi
\providecommand\bibfield[2]{#2}
\providecommand\bibinfo[2]{#2}
\providecommand\natexlab[1]{#1}
\providecommand\showeprint[2][]{arXiv:#2}

\bibitem[\protect\citeauthoryear{Amsaleg and J{\'e}gou}{Amsaleg and
  J{\'e}gou}{2010}]%
        {Amsaleg2010Texmex}
\bibfield{author}{\bibinfo{person}{Laurent Amsaleg} {and}
  \bibinfo{person}{Herv{\'e} J{\'e}gou}.} \bibinfo{year}{2010}\natexlab{}.
\newblock \bibinfo{title}{Datasets for approximate nearest neighbor search}.
\newblock \bibinfo{howpublished}{\url{http://corpus-texmex.irisa.fr/}}.
\newblock
\newblock
\shownote{[Online; accessed 10-July-2024].}


\bibitem[\protect\citeauthoryear{An, Cheng, Park, and Jiang}{An
  et~al\mbox{.}}{2025}]%
        {an2025hyperrag}
\bibfield{author}{\bibinfo{person}{Yuwei An}, \bibinfo{person}{Yihua Cheng},
  \bibinfo{person}{Seo~Jin Park}, {and} \bibinfo{person}{Junchen Jiang}.}
  \bibinfo{year}{2025}\natexlab{}.
\newblock \showarticletitle{Hyperrag: Enhancing quality-efficiency tradeoffs in
  retrieval-augmented generation with reranker kv-cache reuse}.
\newblock \bibinfo{journal}{\emph{arXiv preprint arXiv:2504.02921}}
  (\bibinfo{year}{2025}).
\newblock


\bibitem[\protect\citeauthoryear{Andoni and Indyk}{Andoni and Indyk}{2008}]%
        {andoni2008near}
\bibfield{author}{\bibinfo{person}{Alexandr Andoni} {and}
  \bibinfo{person}{Piotr Indyk}.} \bibinfo{year}{2008}\natexlab{}.
\newblock \showarticletitle{Near-optimal hashing algorithms for approximate
  nearest neighbor in high dimensions}.
\newblock \bibinfo{journal}{\emph{Commun. ACM}} \bibinfo{volume}{51},
  \bibinfo{number}{1} (\bibinfo{year}{2008}), \bibinfo{pages}{117--122}.
\newblock


\bibitem[\protect\citeauthoryear{Andoni, Indyk, Laarhoven, Razenshteyn, and
  Schmidt}{Andoni et~al\mbox{.}}{2015}]%
        {andoni2015practical}
\bibfield{author}{\bibinfo{person}{Alexandr Andoni}, \bibinfo{person}{Piotr
  Indyk}, \bibinfo{person}{Thijs Laarhoven}, \bibinfo{person}{Ilya
  Razenshteyn}, {and} \bibinfo{person}{Ludwig Schmidt}.}
  \bibinfo{year}{2015}\natexlab{}.
\newblock \showarticletitle{Practical and optimal LSH for angular distance}.
\newblock \bibinfo{journal}{\emph{Advances in neural information processing
  systems}}  \bibinfo{volume}{28} (\bibinfo{year}{2015}).
\newblock


\bibitem[\protect\citeauthoryear{Arthur and Vassilvitskii}{Arthur and
  Vassilvitskii}{2006}]%
        {arthur2006k}
\bibfield{author}{\bibinfo{person}{David Arthur} {and} \bibinfo{person}{Sergei
  Vassilvitskii}.} \bibinfo{year}{2006}\natexlab{}.
\newblock \bibinfo{booktitle}{\emph{k-means++: The advantages of careful
  seeding}}.
\newblock \bibinfo{type}{{T}echnical {R}eport}.
  \bibinfo{institution}{Stanford}.
\newblock


\bibitem[\protect\citeauthoryear{Aumüller, Bernhardsson, and
  Faithfull}{Aumüller et~al\mbox{.}}{2018}]%
        {ann-bench:2018}
\bibfield{author}{\bibinfo{person}{Martin Aumüller}, \bibinfo{person}{Erik
  Bernhardsson}, {and} \bibinfo{person}{Alexander Faithfull}.}
  \bibinfo{year}{2018}\natexlab{}.
\newblock \bibinfo{title}{ANN-Benchmarks: A Benchmarking Tool for Approximate
  Nearest Neighbor Algorithms}.
\newblock
\newblock
\showeprint[arxiv]{1807.05614}~[cs.IR]
\urldef\tempurl%
\url{https://arxiv.org/abs/1807.05614}
\showURL{%
\tempurl}


\bibitem[\protect\citeauthoryear{Babenko and Lempitsky}{Babenko and
  Lempitsky}{2015}]%
        {inverted_index}
\bibfield{author}{\bibinfo{person}{Artem Babenko} {and} \bibinfo{person}{Victor
  Lempitsky}.} \bibinfo{year}{2015}\natexlab{}.
\newblock \showarticletitle{The Inverted Multi-Index}.
\newblock \bibinfo{journal}{\emph{IEEE Transactions on Pattern Analysis and
  Machine Intelligence}} \bibinfo{volume}{37}, \bibinfo{number}{6}
  (\bibinfo{year}{2015}), \bibinfo{pages}{1247--1260}.
\newblock
\urldef\tempurl%
\url{https://doi.org/10.1109/TPAMI.2014.2361319}
\showDOI{\tempurl}


\bibitem[\protect\citeauthoryear{Baranchuk, Babenko, and Malkov}{Baranchuk
  et~al\mbox{.}}{2018}]%
        {baranchuk2018revisiting}
\bibfield{author}{\bibinfo{person}{Dmitry Baranchuk}, \bibinfo{person}{Artem
  Babenko}, {and} \bibinfo{person}{Yury Malkov}.}
  \bibinfo{year}{2018}\natexlab{}.
\newblock \showarticletitle{Revisiting the inverted indices for billion-scale
  approximate nearest neighbors}. In \bibinfo{booktitle}{\emph{Proceedings of
  the European Conference on Computer Vision (ECCV)}}.
  \bibinfo{pages}{202--216}.
\newblock


\bibitem[\protect\citeauthoryear{Bordawekar and Shmueli}{Bordawekar and
  Shmueli}{2016}]%
        {bordawekar:corr-abs-1603-07185}
\bibfield{author}{\bibinfo{person}{Rajesh Bordawekar} {and}
  \bibinfo{person}{Oded Shmueli}.} \bibinfo{year}{2016}\natexlab{}.
\newblock \showarticletitle{Enabling Cognitive Intelligence Queries in
  Relational Databases using Low-dimensional Word Embeddings}.
\newblock \bibinfo{journal}{\emph{CoRR}}  \bibinfo{volume}{abs/1603.07185}
  (\bibinfo{year}{2016}).
\newblock
\urldef\tempurl%
\url{http://arxiv.org/abs/1603.07185}
\showURL{%
\tempurl}


\bibitem[\protect\citeauthoryear{Bordawekar and Shmueli}{Bordawekar and
  Shmueli}{2017}]%
        {bordawekar:deem2017}
\bibfield{author}{\bibinfo{person}{Rajesh Bordawekar} {and}
  \bibinfo{person}{Oded Shmueli}.} \bibinfo{year}{2017}\natexlab{}.
\newblock \showarticletitle{Using Word Embedding to Enable Semantic Queries in
  Relational Databases}. In \bibinfo{booktitle}{\emph{Proceedings of the 1st
  Workshop on Data Management for End-to-End Machine Learning}} (Chicago, IL,
  USA) \emph{(\bibinfo{series}{DEEM'17})}. \bibinfo{publisher}{ACM},
  \bibinfo{address}{New York, NY, USA}, Article \bibinfo{articleno}{5},
  \bibinfo{numpages}{4}~pages.
\newblock
\showISBNx{978-1-4503-5026-6}
\urldef\tempurl%
\url{https://doi.org/10.1145/3076246.3076251}
\showDOI{\tempurl}


\bibitem[\protect\citeauthoryear{Charikar}{Charikar}{2002}]%
        {srp-charikar}
\bibfield{author}{\bibinfo{person}{Moses~S. Charikar}.}
  \bibinfo{year}{2002}\natexlab{}.
\newblock \showarticletitle{Similarity Estimation Techniques from Rounding
  Algorithms}. In \bibinfo{booktitle}{\emph{Proceedings of the Thiry-fourth
  Annual ACM Symposium on Theory of Computing}}. \bibinfo{pages}{380--388}.
\newblock


\bibitem[\protect\citeauthoryear{Chen, Xiao, Zhang, Luo, Lian, and Liu}{Chen
  et~al\mbox{.}}{2024}]%
        {chen2024bge}
\bibfield{author}{\bibinfo{person}{Jianlv Chen}, \bibinfo{person}{Shitao Xiao},
  \bibinfo{person}{Peitian Zhang}, \bibinfo{person}{Kun Luo},
  \bibinfo{person}{Defu Lian}, {and} \bibinfo{person}{Zheng Liu}.}
  \bibinfo{year}{2024}\natexlab{}.
\newblock \showarticletitle{Bge m3-embedding: Multi-lingual,
  multi-functionality, multi-granularity text embeddings through self-knowledge
  distillation}.
\newblock \bibinfo{journal}{\emph{arXiv preprint arXiv:2402.03216}}
  (\bibinfo{year}{2024}).
\newblock


\bibitem[\protect\citeauthoryear{Chen, Wang, Li, Ren, Li, Zhu, Li, Liu, Zhang,
  and Wang}{Chen et~al\mbox{.}}{2018}]%
        {ChenW18}
\bibfield{author}{\bibinfo{person}{Qi Chen}, \bibinfo{person}{Haidong Wang},
  \bibinfo{person}{Mingqin Li}, \bibinfo{person}{Gang Ren},
  \bibinfo{person}{Scarlett Li}, \bibinfo{person}{Jeffery Zhu},
  \bibinfo{person}{Jason Li}, \bibinfo{person}{Chuanjie Liu},
  \bibinfo{person}{Lintao Zhang}, {and} \bibinfo{person}{Jingdong Wang}.}
  \bibinfo{year}{2018}\natexlab{}.
\newblock \bibinfo{booktitle}{\emph{SPTAG: A library for fast approximate
  nearest neighbor search}}.
\newblock
\urldef\tempurl%
\url{https://github.com/Microsoft/SPTAG}
\showURL{%
\tempurl}


\bibitem[\protect\citeauthoryear{Chen, Zhao, Wang, Li, Liu, Li, Yang, and
  Wang}{Chen et~al\mbox{.}}{2021}]%
        {chen2021spann}
\bibfield{author}{\bibinfo{person}{Qi Chen}, \bibinfo{person}{Bing Zhao},
  \bibinfo{person}{Haidong Wang}, \bibinfo{person}{Mingqin Li},
  \bibinfo{person}{Chuanjie Liu}, \bibinfo{person}{Zengzhong Li},
  \bibinfo{person}{Mao Yang}, {and} \bibinfo{person}{Jingdong Wang}.}
  \bibinfo{year}{2021}\natexlab{}.
\newblock \showarticletitle{Spann: Highly-efficient billion-scale approximate
  nearest neighborhood search}.
\newblock \bibinfo{journal}{\emph{Advances in Neural Information Processing
  Systems}}  \bibinfo{volume}{34} (\bibinfo{year}{2021}),
  \bibinfo{pages}{5199--5212}.
\newblock


\bibitem[\protect\citeauthoryear{Chroma}{Chroma}{2022}]%
        {chroma}
\bibfield{author}{\bibinfo{person}{Chroma}.} \bibinfo{year}{2022}\natexlab{}.
\newblock \bibinfo{title}{Chroma}.
\newblock \bibinfo{howpublished}{\url{https://github.com/chroma-core/chroma}}.
\newblock


\bibitem[\protect\citeauthoryear{Ciaccia, Patella, Zezula,
  et~al\mbox{.}}{Ciaccia et~al\mbox{.}}{1997}]%
        {ciaccia1997m}
\bibfield{author}{\bibinfo{person}{Paolo Ciaccia}, \bibinfo{person}{Marco
  Patella}, \bibinfo{person}{Pavel Zezula}, {et~al\mbox{.}}}
  \bibinfo{year}{1997}\natexlab{}.
\newblock \showarticletitle{M-tree: An efficient access method for similarity
  search in metric spaces}. In \bibinfo{booktitle}{\emph{Vldb}},
  Vol.~\bibinfo{volume}{97}. Citeseer, \bibinfo{pages}{426--435}.
\newblock


\bibitem[\protect\citeauthoryear{Datar, Immorlica, Indyk, and Mirrokni}{Datar
  et~al\mbox{.}}{2004}]%
        {datar2004locality}
\bibfield{author}{\bibinfo{person}{Mayur Datar}, \bibinfo{person}{Nicole
  Immorlica}, \bibinfo{person}{Piotr Indyk}, {and} \bibinfo{person}{Vahab~S
  Mirrokni}.} \bibinfo{year}{2004}\natexlab{}.
\newblock \showarticletitle{Locality-sensitive hashing scheme based on p-stable
  distributions}. In \bibinfo{booktitle}{\emph{Proceedings of the twentieth
  annual symposium on Computational geometry}}. \bibinfo{pages}{253--262}.
\newblock


\bibitem[\protect\citeauthoryear{Douze, Guzhva, Deng, Johnson, Szilvasy,
  Mazaré, Lomeli, Hosseini, and Jégou}{Douze et~al\mbox{.}}{2024}]%
        {douze2024faiss}
\bibfield{author}{\bibinfo{person}{Matthijs Douze}, \bibinfo{person}{Alexandr
  Guzhva}, \bibinfo{person}{Chengqi Deng}, \bibinfo{person}{Jeff Johnson},
  \bibinfo{person}{Gergely Szilvasy}, \bibinfo{person}{Pierre-Emmanuel
  Mazaré}, \bibinfo{person}{Maria Lomeli}, \bibinfo{person}{Lucas Hosseini},
  {and} \bibinfo{person}{Hervé Jégou}.} \bibinfo{year}{2024}\natexlab{}.
\newblock \showarticletitle{The Faiss library}.
\newblock  (\bibinfo{year}{2024}).
\newblock
\showeprint[arxiv]{2401.08281}~[cs.LG]


\bibitem[\protect\citeauthoryear{Elasticsearch}{Elasticsearch}{2015}]%
        {elastic}
\bibfield{author}{\bibinfo{person}{Elasticsearch}.}
  \bibinfo{year}{2015}\natexlab{}.
\newblock \bibinfo{title}{Elasticsearch}.
\newblock \bibinfo{howpublished}{\url{https://www.elastic.co/}}.
\newblock


\bibitem[\protect\citeauthoryear{Friedman, Bentley, and Finkel}{Friedman
  et~al\mbox{.}}{1977}]%
        {friedman1977algorithm}
\bibfield{author}{\bibinfo{person}{Jerome~H Friedman},
  \bibinfo{person}{Jon~Louis Bentley}, {and} \bibinfo{person}{Raphael~Ari
  Finkel}.} \bibinfo{year}{1977}\natexlab{}.
\newblock \showarticletitle{An algorithm for finding best matches in
  logarithmic expected time}.
\newblock \bibinfo{journal}{\emph{ACM Transactions on Mathematical Software
  (TOMS)}} \bibinfo{volume}{3}, \bibinfo{number}{3} (\bibinfo{year}{1977}),
  \bibinfo{pages}{209--226}.
\newblock


\bibitem[\protect\citeauthoryear{Gao and Callan}{Gao and Callan}{2022}]%
        {rerankGAO}
\bibfield{author}{\bibinfo{person}{Luyu Gao} {and} \bibinfo{person}{Jamie
  Callan}.} \bibinfo{year}{2022}\natexlab{}.
\newblock \showarticletitle{Long Document Re-ranking with Modular Re-ranker}.
  In \bibinfo{booktitle}{\emph{Proceedings of the 45th International ACM SIGIR
  Conference on Research and Development in Information Retrieval}} (Madrid,
  Spain) \emph{(\bibinfo{series}{SIGIR '22})}. \bibinfo{publisher}{Association
  for Computing Machinery}, \bibinfo{address}{New York, NY, USA},
  \bibinfo{pages}{2371–2376}.
\newblock
\showISBNx{9781450387323}
\urldef\tempurl%
\url{https://doi.org/10.1145/3477495.3531860}
\showDOI{\tempurl}


\bibitem[\protect\citeauthoryear{Gollapudi, Karia, Sivashankar, Krishnaswamy,
  Begwani, Raz, Lin, Zhang, Mahapatro, Srinivasan, et~al\mbox{.}}{Gollapudi
  et~al\mbox{.}}{2023}]%
        {gollapudi2023filtered}
\bibfield{author}{\bibinfo{person}{Siddharth Gollapudi}, \bibinfo{person}{Neel
  Karia}, \bibinfo{person}{Varun Sivashankar}, \bibinfo{person}{Ravishankar
  Krishnaswamy}, \bibinfo{person}{Nikit Begwani}, \bibinfo{person}{Swapnil
  Raz}, \bibinfo{person}{Yiyong Lin}, \bibinfo{person}{Yin Zhang},
  \bibinfo{person}{Neelam Mahapatro}, \bibinfo{person}{Premkumar Srinivasan},
  {et~al\mbox{.}}} \bibinfo{year}{2023}\natexlab{}.
\newblock \showarticletitle{Filtered-diskann: Graph algorithms for approximate
  nearest neighbor search with filters}. In
  \bibinfo{booktitle}{\emph{Proceedings of the ACM Web Conference 2023}}.
  \bibinfo{pages}{3406--3416}.
\newblock


\bibitem[\protect\citeauthoryear{Gong, Zeng, and Chen}{Gong
  et~al\mbox{.}}{2025}]%
        {gong:shortcut}
\bibfield{author}{\bibinfo{person}{Zengyang Gong}, \bibinfo{person}{Yuxiang
  Zeng}, {and} \bibinfo{person}{Lei Chen}.} \bibinfo{year}{2025}\natexlab{}.
\newblock \showarticletitle{Accelerating Approximate Nearest Neighbor Search in
  Hierarchical Graphs: Efficient Level Navigation with Shortcuts}.
\newblock \bibinfo{journal}{\emph{Proceedings of the VLDB Endowment}}
  \bibinfo{volume}{18}, \bibinfo{number}{10} (\bibinfo{year}{2025}).
\newblock


\bibitem[\protect\citeauthoryear{Guo, Sun, Lindgren, Geng, Simcha, Chern, and
  Kumar}{Guo et~al\mbox{.}}{2020}]%
        {guo2020accelerating}
\bibfield{author}{\bibinfo{person}{Ruiqi Guo}, \bibinfo{person}{Philip Sun},
  \bibinfo{person}{Erik Lindgren}, \bibinfo{person}{Quan Geng},
  \bibinfo{person}{David Simcha}, \bibinfo{person}{Felix Chern}, {and}
  \bibinfo{person}{Sanjiv Kumar}.} \bibinfo{year}{2020}\natexlab{}.
\newblock \showarticletitle{Accelerating large-scale inference with anisotropic
  vector quantization}. In \bibinfo{booktitle}{\emph{International Conference
  on Machine Learning}}. PMLR, \bibinfo{pages}{3887--3896}.
\newblock


\bibitem[\protect\citeauthoryear{Guttman}{Guttman}{1984}]%
        {guttman1984r}
\bibfield{author}{\bibinfo{person}{Antonin Guttman}.}
  \bibinfo{year}{1984}\natexlab{}.
\newblock \showarticletitle{R-trees: A dynamic index structure for spatial
  searching}. In \bibinfo{booktitle}{\emph{Proceedings of the 1984 ACM SIGMOD
  international conference on Management of data}}. \bibinfo{pages}{47--57}.
\newblock


\bibitem[\protect\citeauthoryear{Han, Liu, and Wang}{Han et~al\mbox{.}}{2023}]%
        {han2023comprehensive}
\bibfield{author}{\bibinfo{person}{Yikun Han}, \bibinfo{person}{Chunjiang Liu},
  {and} \bibinfo{person}{Pengfei Wang}.} \bibinfo{year}{2023}\natexlab{}.
\newblock \showarticletitle{A comprehensive survey on vector database: Storage
  and retrieval technique, challenge}.
\newblock \bibinfo{journal}{\emph{arXiv preprint arXiv:2310.11703}}
  (\bibinfo{year}{2023}).
\newblock


\bibitem[\protect\citeauthoryear{{IBM Corp.}}{{IBM Corp.}}{2025a}]%
        {db2vectors}
\bibfield{author}{\bibinfo{person}{{IBM Corp.}}}
  \bibinfo{year}{2025}\natexlab{a}.
\newblock \bibinfo{title}{{Db2 for Linux, UNIX and Windows Vector values
  12.1.x}}.
\newblock
\newblock
\urldef\tempurl%
\url{https://www.ibm.com/docs/en/db2/12.1.x?topic=list-vector-values}
\showURL{%
\tempurl}


\bibitem[\protect\citeauthoryear{{IBM Corp.}}{{IBM Corp.}}{2025b}]%
        {sqldi}
\bibfield{author}{\bibinfo{person}{{IBM Corp.}}}
  \bibinfo{year}{2025}\natexlab{b}.
\newblock \bibinfo{title}{{Running AI Queries with SQL Data Insights}}.
\newblock
\newblock
\urldef\tempurl%
\url{https://www.ibm.com/docs/en/db2-for-zos/13.0.0?topic=running-ai-queries-sql-data-insights}
\showURL{%
\tempurl}


\bibitem[\protect\citeauthoryear{Jafari, Maurya, Nagarkar, Islam, and
  Crushev}{Jafari et~al\mbox{.}}{2021}]%
        {jafari2021survey}
\bibfield{author}{\bibinfo{person}{Omid Jafari}, \bibinfo{person}{Preeti
  Maurya}, \bibinfo{person}{Parth Nagarkar},
  \bibinfo{person}{Khandker~Mushfiqul Islam}, {and}
  \bibinfo{person}{Chidambaram Crushev}.} \bibinfo{year}{2021}\natexlab{}.
\newblock \showarticletitle{A survey on locality sensitive hashing algorithms
  and their applications}.
\newblock \bibinfo{journal}{\emph{arXiv preprint arXiv:2102.08942}}
  (\bibinfo{year}{2021}).
\newblock


\bibitem[\protect\citeauthoryear{Jayaram~Subramanya, Devvrit, Simhadri,
  Krishnawamy, and Kadekodi}{Jayaram~Subramanya et~al\mbox{.}}{2019}]%
        {jayaram2019diskann}
\bibfield{author}{\bibinfo{person}{Suhas Jayaram~Subramanya},
  \bibinfo{person}{Fnu Devvrit}, \bibinfo{person}{Harsha~Vardhan Simhadri},
  \bibinfo{person}{Ravishankar Krishnawamy}, {and} \bibinfo{person}{Rohan
  Kadekodi}.} \bibinfo{year}{2019}\natexlab{}.
\newblock \showarticletitle{Diskann: Fast accurate billion-point nearest
  neighbor search on a single node}.
\newblock \bibinfo{journal}{\emph{Advances in neural information processing
  Systems}}  \bibinfo{volume}{32} (\bibinfo{year}{2019}).
\newblock


\bibitem[\protect\citeauthoryear{J{\'{e}}gou, Tavenard, Douze, and
  Amsaleg}{J{\'{e}}gou et~al\mbox{.}}{2011}]%
        {source_coding}
\bibfield{author}{\bibinfo{person}{Herv{\'{e}} J{\'{e}}gou},
  \bibinfo{person}{Romain Tavenard}, \bibinfo{person}{Matthijs Douze}, {and}
  \bibinfo{person}{Laurent Amsaleg}.} \bibinfo{year}{2011}\natexlab{}.
\newblock \showarticletitle{Searching in one billion vectors: re-rank with
  source coding}.
\newblock \bibinfo{journal}{\emph{CoRR}}  \bibinfo{volume}{abs/1102.3828}
  (\bibinfo{year}{2011}).
\newblock
\showeprint[arXiv]{1102.3828}
\urldef\tempurl%
\url{http://arxiv.org/abs/1102.3828}
\showURL{%
\tempurl}


\bibitem[\protect\citeauthoryear{Johnson, Douze, and J{\'e}gou}{Johnson
  et~al\mbox{.}}{2019}]%
        {johnson2019billion}
\bibfield{author}{\bibinfo{person}{Jeff Johnson}, \bibinfo{person}{Matthijs
  Douze}, {and} \bibinfo{person}{Herv{\'e} J{\'e}gou}.}
  \bibinfo{year}{2019}\natexlab{}.
\newblock \showarticletitle{Billion-scale similarity search with GPUs}.
\newblock \bibinfo{journal}{\emph{IEEE Transactions on Big Data}}
  \bibinfo{volume}{7}, \bibinfo{number}{3} (\bibinfo{year}{2019}),
  \bibinfo{pages}{535--547}.
\newblock


\bibitem[\protect\citeauthoryear{Jégou, Douze, and Schmid}{Jégou
  et~al\mbox{.}}{2011}]%
        {product_quantization}
\bibfield{author}{\bibinfo{person}{Herve Jégou}, \bibinfo{person}{Matthijs
  Douze}, {and} \bibinfo{person}{Cordelia Schmid}.}
  \bibinfo{year}{2011}\natexlab{}.
\newblock \showarticletitle{Product Quantization for Nearest Neighbor Search}.
\newblock \bibinfo{journal}{\emph{IEEE Transactions on Pattern Analysis and
  Machine Intelligence}} \bibinfo{volume}{33}, \bibinfo{number}{1}
  (\bibinfo{year}{2011}), \bibinfo{pages}{117--128}.
\newblock
\urldef\tempurl%
\url{https://doi.org/10.1109/TPAMI.2010.57}
\showDOI{\tempurl}


\bibitem[\protect\citeauthoryear{Kalantidis and Avrithis}{Kalantidis and
  Avrithis}{2014}]%
        {kalantidis2014locally}
\bibfield{author}{\bibinfo{person}{Yannis Kalantidis} {and}
  \bibinfo{person}{Yannis Avrithis}.} \bibinfo{year}{2014}\natexlab{}.
\newblock \showarticletitle{Locally optimized product quantization for
  approximate nearest neighbor search}. In
  \bibinfo{booktitle}{\emph{Proceedings of the IEEE conference on computer
  vision and pattern recognition}}. \bibinfo{pages}{2321--2328}.
\newblock


\bibitem[\protect\citeauthoryear{Katsis, Rosenthal, Fadnis, Gunasekara, Lee,
  Popa, Shah, Zhu, Contractor, and Danilevsky}{Katsis et~al\mbox{.}}{2025}]%
        {katsis2025mtrag}
\bibfield{author}{\bibinfo{person}{Yannis Katsis}, \bibinfo{person}{Sara
  Rosenthal}, \bibinfo{person}{Kshitij Fadnis}, \bibinfo{person}{Chulaka
  Gunasekara}, \bibinfo{person}{Young-Suk Lee}, \bibinfo{person}{Lucian Popa},
  \bibinfo{person}{Vraj Shah}, \bibinfo{person}{Huaiyu Zhu},
  \bibinfo{person}{Danish Contractor}, {and} \bibinfo{person}{Marina
  Danilevsky}.} \bibinfo{year}{2025}\natexlab{}.
\newblock \showarticletitle{Mtrag: A multi-turn conversational benchmark for
  evaluating retrieval-augmented generation systems}.
\newblock \bibinfo{journal}{\emph{Transactions of the Association for
  Computational Linguistics}}  \bibinfo{volume}{13} (\bibinfo{year}{2025}),
  \bibinfo{pages}{784--808}.
\newblock


\bibitem[\protect\citeauthoryear{Lewis, Perez, Piktus, Petroni, Karpukhin,
  Goyal, K{\"u}ttler, Lewis, Yih, Rockt{\"a}schel, et~al\mbox{.}}{Lewis
  et~al\mbox{.}}{2020}]%
        {lewis2020retrieval}
\bibfield{author}{\bibinfo{person}{Patrick Lewis}, \bibinfo{person}{Ethan
  Perez}, \bibinfo{person}{Aleksandra Piktus}, \bibinfo{person}{Fabio Petroni},
  \bibinfo{person}{Vladimir Karpukhin}, \bibinfo{person}{Naman Goyal},
  \bibinfo{person}{Heinrich K{\"u}ttler}, \bibinfo{person}{Mike Lewis},
  \bibinfo{person}{Wen-tau Yih}, \bibinfo{person}{Tim Rockt{\"a}schel},
  {et~al\mbox{.}}} \bibinfo{year}{2020}\natexlab{}.
\newblock \showarticletitle{Retrieval-augmented generation for
  knowledge-intensive nlp tasks}.
\newblock \bibinfo{journal}{\emph{Advances in Neural Information Processing
  Systems}}  \bibinfo{volume}{33} (\bibinfo{year}{2020}),
  \bibinfo{pages}{9459--9474}.
\newblock


\bibitem[\protect\citeauthoryear{Malkov et~al\mbox{.}}{Malkov
  et~al\mbox{.}}{[n.d.]}]%
        {malkov2018hnswlib}
\bibfield{author}{\bibinfo{person}{Yury Malkov} {et~al\mbox{.}}}
  \bibinfo{year}{[n.d.]}\natexlab{}.
\newblock \bibinfo{booktitle}{\emph{hnswlib: Hierarchical Navigable Small World
  graphs}}.
\newblock
\urldef\tempurl%
\url{https://github.com/nmslib/hnswlib}
\showURL{%
\tempurl}
\newblock
\shownote{[Online; accessed 28-September-2025].}


\bibitem[\protect\citeauthoryear{Malkov, Ponomarenko, Logvinov, and
  Krylov}{Malkov et~al\mbox{.}}{2014}]%
        {malkov2014approximate}
\bibfield{author}{\bibinfo{person}{Yury Malkov}, \bibinfo{person}{Alexander
  Ponomarenko}, \bibinfo{person}{Andrey Logvinov}, {and}
  \bibinfo{person}{Vladimir Krylov}.} \bibinfo{year}{2014}\natexlab{}.
\newblock \showarticletitle{Approximate nearest neighbor algorithm based on
  navigable small world graphs}.
\newblock \bibinfo{journal}{\emph{Information Systems}}  \bibinfo{volume}{45}
  (\bibinfo{year}{2014}), \bibinfo{pages}{61--68}.
\newblock


\bibitem[\protect\citeauthoryear{Malkov and Yashunin}{Malkov and
  Yashunin}{2018}]%
        {malkov2018efficient}
\bibfield{author}{\bibinfo{person}{Yu~A Malkov} {and} \bibinfo{person}{Dmitry~A
  Yashunin}.} \bibinfo{year}{2018}\natexlab{}.
\newblock \showarticletitle{Efficient and robust approximate nearest neighbor
  search using hierarchical navigable small world graphs}.
\newblock \bibinfo{journal}{\emph{IEEE transactions on pattern analysis and
  machine intelligence}} \bibinfo{volume}{42}, \bibinfo{number}{4}
  (\bibinfo{year}{2018}), \bibinfo{pages}{824--836}.
\newblock


\bibitem[\protect\citeauthoryear{Manohar, Shen, Blelloch, Dhulipala, Gu,
  Simhadri, and Sun}{Manohar et~al\mbox{.}}{2024}]%
        {manohar2024parlayann}
\bibfield{author}{\bibinfo{person}{Magdalen~Dobson Manohar},
  \bibinfo{person}{Zheqi Shen}, \bibinfo{person}{Guy Blelloch},
  \bibinfo{person}{Laxman Dhulipala}, \bibinfo{person}{Yan Gu},
  \bibinfo{person}{Harsha~Vardhan Simhadri}, {and} \bibinfo{person}{Yihan
  Sun}.} \bibinfo{year}{2024}\natexlab{}.
\newblock \showarticletitle{ParlayANN: Scalable and Deterministic Parallel
  Graph-Based Approximate Nearest Neighbor Search Algorithms}. In
  \bibinfo{booktitle}{\emph{Proceedings of the 29th ACM SIGPLAN Annual
  Symposium on Principles and Practice of Parallel Programming}}.
  \bibinfo{pages}{270--285}.
\newblock


\bibitem[\protect\citeauthoryear{Mikolov, Chen, Corrado, and Dean}{Mikolov
  et~al\mbox{.}}{2013}]%
        {mikolov:corr-abs-1301-3781}
\bibfield{author}{\bibinfo{person}{Tomas Mikolov}, \bibinfo{person}{Kai Chen},
  \bibinfo{person}{Greg Corrado}, {and} \bibinfo{person}{Jeffrey Dean}.}
  \bibinfo{year}{2013}\natexlab{}.
\newblock \showarticletitle{Efficient Estimation of Word Representations in
  Vector Space}.
\newblock \bibinfo{journal}{\emph{CoRR}}  \bibinfo{volume}{abs/1301.3781}
  (\bibinfo{year}{2013}).
\newblock
\urldef\tempurl%
\url{http://arxiv.org/abs/1301.3781}
\showURL{%
\tempurl}


\bibitem[\protect\citeauthoryear{Muja and Lowe}{Muja and Lowe}{2014}]%
        {muja2014scalable}
\bibfield{author}{\bibinfo{person}{Marius Muja} {and} \bibinfo{person}{David~G
  Lowe}.} \bibinfo{year}{2014}\natexlab{}.
\newblock \showarticletitle{Scalable nearest neighbor algorithms for high
  dimensional data}.
\newblock \bibinfo{journal}{\emph{IEEE transactions on pattern analysis and
  machine intelligence}} \bibinfo{volume}{36}, \bibinfo{number}{11}
  (\bibinfo{year}{2014}), \bibinfo{pages}{2227--2240}.
\newblock


\bibitem[\protect\citeauthoryear{Munoz, Gon{\c{c}}alves, Dias, and
  Torres}{Munoz et~al\mbox{.}}{2019}]%
        {munoz2019hierarchical}
\bibfield{author}{\bibinfo{person}{Javier~Vargas Munoz},
  \bibinfo{person}{Marcos~A Gon{\c{c}}alves}, \bibinfo{person}{Zanoni Dias},
  {and} \bibinfo{person}{Ricardo da~S Torres}.}
  \bibinfo{year}{2019}\natexlab{}.
\newblock \showarticletitle{Hierarchical clustering-based graphs for large
  scale approximate nearest neighbor search}.
\newblock \bibinfo{journal}{\emph{Pattern Recognition}}  \bibinfo{volume}{96}
  (\bibinfo{year}{2019}), \bibinfo{pages}{106970}.
\newblock


\bibitem[\protect\citeauthoryear{{Nvidia Inc.}}{{Nvidia Inc.}}{2024}]%
        {cuvs}
\bibfield{author}{\bibinfo{person}{{Nvidia Inc.}}}
  \bibinfo{year}{2024}\natexlab{}.
\newblock \bibinfo{title}{{cuVS: Vector Search and Clustering on the GPU}}.
\newblock
\newblock
\urldef\tempurl%
\url{https://github.com/rapidsai/cuvs}
\showURL{%
\tempurl}


\bibitem[\protect\citeauthoryear{Ootomo, Naruse, Nolet, Wang, Feher, and
  Wang}{Ootomo et~al\mbox{.}}{2023}]%
        {ootomo2023cagra}
\bibfield{author}{\bibinfo{person}{Hiroyuki Ootomo}, \bibinfo{person}{Akira
  Naruse}, \bibinfo{person}{Corey Nolet}, \bibinfo{person}{Ray Wang},
  \bibinfo{person}{Tamas Feher}, {and} \bibinfo{person}{Yong Wang}.}
  \bibinfo{year}{2023}\natexlab{}.
\newblock \showarticletitle{Cagra: Highly parallel graph construction and
  approximate nearest neighbor search for gpus}.
\newblock \bibinfo{journal}{\emph{arXiv preprint arXiv:2308.15136}}
  (\bibinfo{year}{2023}).
\newblock


\bibitem[\protect\citeauthoryear{{Oracle Inc.}}{{Oracle Inc.}}{2025}]%
        {oracle26i}
\bibfield{author}{\bibinfo{person}{{Oracle Inc.}}}
  \bibinfo{year}{2025}\natexlab{}.
\newblock \bibinfo{title}{{Oracle AI Database 26i}}.
\newblock
\newblock
\urldef\tempurl%
\url{https://www.oracle.com/database/ai-native-database-26ai}
\showURL{%
\tempurl}


\bibitem[\protect\citeauthoryear{Pan, Wang, and Li}{Pan et~al\mbox{.}}{2023}]%
        {pan2023survey}
\bibfield{author}{\bibinfo{person}{James~Jie Pan}, \bibinfo{person}{Jianguo
  Wang}, {and} \bibinfo{person}{Guoliang Li}.} \bibinfo{year}{2023}\natexlab{}.
\newblock \showarticletitle{Survey of vector database management systems}.
\newblock \bibinfo{journal}{\emph{arXiv preprint arXiv:2310.14021}}
  (\bibinfo{year}{2023}).
\newblock


\bibitem[\protect\citeauthoryear{Pennington, Socher, and Manning}{Pennington
  et~al\mbox{.}}{2014}]%
        {pennington2014glove}
\bibfield{author}{\bibinfo{person}{Jeffrey Pennington},
  \bibinfo{person}{Richard Socher}, {and} \bibinfo{person}{Christopher~D
  Manning}.} \bibinfo{year}{2014}\natexlab{}.
\newblock \showarticletitle{Glove: Global vectors for word representation}. In
  \bibinfo{booktitle}{\emph{Proceedings of the 2014 conference on empirical
  methods in natural language processing (EMNLP)}}.
  \bibinfo{pages}{1532--1543}.
\newblock


\bibitem[\protect\citeauthoryear{Ren, Zhang, and Li}{Ren et~al\mbox{.}}{2020}]%
        {ren2020hm}
\bibfield{author}{\bibinfo{person}{Jie Ren}, \bibinfo{person}{Minjia Zhang},
  {and} \bibinfo{person}{Dong Li}.} \bibinfo{year}{2020}\natexlab{}.
\newblock \showarticletitle{Hm-ann: Efficient billion-point nearest neighbor
  search on heterogeneous memory}.
\newblock \bibinfo{journal}{\emph{Advances in Neural Information Processing
  Systems}}  \bibinfo{volume}{33} (\bibinfo{year}{2020}),
  \bibinfo{pages}{10672--10684}.
\newblock


\bibitem[\protect\citeauthoryear{Shim, Oh, Roh, Do, and Lee}{Shim
  et~al\mbox{.}}{2025}]%
        {shim:vldb-ssd}
\bibfield{author}{\bibinfo{person}{Joobo Shim}, \bibinfo{person}{Jaewon Oh},
  \bibinfo{person}{HongChan Roh}, \bibinfo{person}{Jaeyoung Do}, {and}
  \bibinfo{person}{Sang-Won Lee}.} \bibinfo{year}{2025}\natexlab{}.
\newblock \showarticletitle{Turbocharging Vector Databases using Modern SSDs}.
\newblock \bibinfo{journal}{\emph{Proceedings of the VLDB Endowment}}
  \bibinfo{volume}{18}, \bibinfo{number}{11} (\bibinfo{year}{2025}).
\newblock


\bibitem[\protect\citeauthoryear{Spotify}{Spotify}{2017}]%
        {annoy}
\bibfield{author}{\bibinfo{person}{Spotify}.} \bibinfo{year}{2017}\natexlab{}.
\newblock \bibinfo{title}{ANNOY}.
\newblock \bibinfo{howpublished}{\url{https://github.com/spotify/annoy}}.
\newblock


\bibitem[\protect\citeauthoryear{Sun, Simcha, Dopson, Guo, and Kumar}{Sun
  et~al\mbox{.}}{2024}]%
        {sun2024soar}
\bibfield{author}{\bibinfo{person}{Philip Sun}, \bibinfo{person}{David Simcha},
  \bibinfo{person}{Dave Dopson}, \bibinfo{person}{Ruiqi Guo}, {and}
  \bibinfo{person}{Sanjiv Kumar}.} \bibinfo{year}{2024}\natexlab{}.
\newblock \showarticletitle{SOAR: improved indexing for approximate nearest
  neighbor search}.
\newblock \bibinfo{journal}{\emph{Advances in Neural Information Processing
  Systems}}  \bibinfo{volume}{36} (\bibinfo{year}{2024}).
\newblock


\bibitem[\protect\citeauthoryear{Vaswani, Shazeer, Parmar, Uszkoreit, Jones,
  Gomez, Kaiser, and Polosukhin}{Vaswani et~al\mbox{.}}{2023}]%
        {vaswani2023attentionneed}
\bibfield{author}{\bibinfo{person}{Ashish Vaswani}, \bibinfo{person}{Noam
  Shazeer}, \bibinfo{person}{Niki Parmar}, \bibinfo{person}{Jakob Uszkoreit},
  \bibinfo{person}{Llion Jones}, \bibinfo{person}{Aidan~N. Gomez},
  \bibinfo{person}{Lukasz Kaiser}, {and} \bibinfo{person}{Illia Polosukhin}.}
  \bibinfo{year}{2023}\natexlab{}.
\newblock \bibinfo{title}{Attention Is All You Need}.
\newblock
\newblock
\showeprint[arxiv]{1706.03762}~[cs.CL]
\urldef\tempurl%
\url{https://arxiv.org/abs/1706.03762}
\showURL{%
\tempurl}


\bibitem[\protect\citeauthoryear{Wang, Yi, Guo, Jin, Xu, Li, Wang, Guo, Li, Xu,
  et~al\mbox{.}}{Wang et~al\mbox{.}}{2021}]%
        {wang2021milvus}
\bibfield{author}{\bibinfo{person}{Jianguo Wang}, \bibinfo{person}{Xiaomeng
  Yi}, \bibinfo{person}{Rentong Guo}, \bibinfo{person}{Hai Jin},
  \bibinfo{person}{Peng Xu}, \bibinfo{person}{Shengjun Li},
  \bibinfo{person}{Xiangyu Wang}, \bibinfo{person}{Xiangzhou Guo},
  \bibinfo{person}{Chengming Li}, \bibinfo{person}{Xiaohai Xu},
  {et~al\mbox{.}}} \bibinfo{year}{2021}\natexlab{}.
\newblock \showarticletitle{Milvus: A purpose-built vector data management
  system}. In \bibinfo{booktitle}{\emph{Proceedings of the 2021 International
  Conference on Management of Data}}. \bibinfo{pages}{2614--2627}.
\newblock


\bibitem[\protect\citeauthoryear{Wang and Zhang}{Wang and Zhang}{2018}]%
        {wang2018composite}
\bibfield{author}{\bibinfo{person}{Jingdong Wang} {and} \bibinfo{person}{Ting
  Zhang}.} \bibinfo{year}{2018}\natexlab{}.
\newblock \showarticletitle{Composite quantization}.
\newblock \bibinfo{journal}{\emph{IEEE transactions on pattern analysis and
  machine intelligence}} \bibinfo{volume}{41}, \bibinfo{number}{6}
  (\bibinfo{year}{2018}), \bibinfo{pages}{1308--1322}.
\newblock


\bibitem[\protect\citeauthoryear{Wang, Zhang, Song, Sebe, and Shen}{Wang
  et~al\mbox{.}}{2016}]%
        {learn_to_hash}
\bibfield{author}{\bibinfo{person}{Jingdong Wang}, \bibinfo{person}{Ting
  Zhang}, \bibinfo{person}{Jingkuan Song}, \bibinfo{person}{Nicu Sebe}, {and}
  \bibinfo{person}{Heng~Tao Shen}.} \bibinfo{year}{2016}\natexlab{}.
\newblock \showarticletitle{A Survey on Learning to Hash}.
\newblock \bibinfo{journal}{\emph{CoRR}}  \bibinfo{volume}{abs/1606.00185}
  (\bibinfo{year}{2016}).
\newblock
\showeprint[arXiv]{1606.00185}
\urldef\tempurl%
\url{http://arxiv.org/abs/1606.00185}
\showURL{%
\tempurl}


\bibitem[\protect\citeauthoryear{Yang, Li, Fang, and Wei}{Yang
  et~al\mbox{.}}{2020}]%
        {yang2020pase}
\bibfield{author}{\bibinfo{person}{Wen Yang}, \bibinfo{person}{Tao Li},
  \bibinfo{person}{Gai Fang}, {and} \bibinfo{person}{Hong Wei}.}
  \bibinfo{year}{2020}\natexlab{}.
\newblock \showarticletitle{Pase: Postgresql ultra-high-dimensional approximate
  nearest neighbor search extension}. In \bibinfo{booktitle}{\emph{Proceedings
  of the 2020 ACM SIGMOD international conference on management of data}}.
  \bibinfo{pages}{2241--2253}.
\newblock


\bibitem[\protect\citeauthoryear{Yue, Zheng, Xu, Xu, Zhang, Du, Gao, Zhou, and
  Jensen}{Yue et~al\mbox{.}}{2025}]%
        {yue:vldb-margo}
\bibfield{author}{\bibinfo{person}{Ziyang Yue}, \bibinfo{person}{Bolong Zheng},
  \bibinfo{person}{Ling Xu}, \bibinfo{person}{Kanru Xu},
  \bibinfo{person}{Shuhao Zhang}, \bibinfo{person}{Yajuan Du},
  \bibinfo{person}{Yunjun Gao}, \bibinfo{person}{Xiaofang Zhou}, {and}
  \bibinfo{person}{Christian Jensen}.} \bibinfo{year}{2025}\natexlab{}.
\newblock \showarticletitle{Select Edges Wisely: Monotonic Path Aware Graph
  Layout Optimization for Disk-based ANN Search}.
\newblock \bibinfo{journal}{\emph{Proceedings of the VLDB Endowment}}
  \bibinfo{volume}{18}, \bibinfo{number}{11} (\bibinfo{year}{2025}).
\newblock


\bibitem[\protect\citeauthoryear{Zhang, Xu, Chen, Sui, Xie, Cai, Chen, He,
  Yang, Yang, et~al\mbox{.}}{Zhang et~al\mbox{.}}{2023}]%
        {zhang2023vbase}
\bibfield{author}{\bibinfo{person}{Qianxi Zhang}, \bibinfo{person}{Shuotao Xu},
  \bibinfo{person}{Qi Chen}, \bibinfo{person}{Guoxin Sui},
  \bibinfo{person}{Jiadong Xie}, \bibinfo{person}{Zhizhen Cai},
  \bibinfo{person}{Yaoqi Chen}, \bibinfo{person}{Yinxuan He},
  \bibinfo{person}{Yuqing Yang}, \bibinfo{person}{Fan Yang}, {et~al\mbox{.}}}
  \bibinfo{year}{2023}\natexlab{}.
\newblock \showarticletitle{$\{$VBASE$\}$: Unifying Online Vector Similarity
  Search and Relational Queries via Relaxed Monotonicity}. In
  \bibinfo{booktitle}{\emph{17th USENIX Symposium on Operating Systems Design
  and Implementation (OSDI 23)}}. \bibinfo{pages}{377--395}.
\newblock


\bibitem[\protect\citeauthoryear{Zhao, Tan, and Li}{Zhao et~al\mbox{.}}{2020}]%
        {zhao2020song}
\bibfield{author}{\bibinfo{person}{Weijie Zhao}, \bibinfo{person}{Shulong Tan},
  {and} \bibinfo{person}{Ping Li}.} \bibinfo{year}{2020}\natexlab{}.
\newblock \showarticletitle{Song: Approximate nearest neighbor search on gpu}.
  In \bibinfo{booktitle}{\emph{2020 IEEE 36th International Conference on Data
  Engineering (ICDE)}}. IEEE, \bibinfo{pages}{1033--1044}.
\newblock


\end{thebibliography}

\end{document}